\documentclass[aps,superscriptaddress,twocolumn,showpacs]{revtex4-1}

\pdfoutput=1

\usepackage{graphicx} % May include graphics.

\usepackage{float} % May put graphics exactly where you want them

\usepackage{subfloat} % May include sub-figures

\usepackage{appendix} % May include appendices.

\usepackage{amsmath} % Nice maths

\usepackage{hyperref} % Makes references and citations clickable in a pdf viewer. Also may include urls in text.

\usepackage{mathtools} % Arrows and other nice things

\usepackage{subfig} % Allows insertion of subfigures

\usepackage{epstopdf} % Name says it all

\usepackage{amsfonts} % Nice Math fonts

\usepackage{bbold}

\usepackage{braket}

\usepackage{program}

\usepackage{sidecap}

\hypersetup{
bookmarks=true,
colorlinks=true,
citecolor = blue,
urlcolor=blue,
pdfstartview={FitH}
}

\newcommand\given[1][]{\:#1\vert\:}

\begin{document}

%\linenumbers

\title{Large deviation analysis of a simple information engine}

\author{Michael Maitland}

\author{Stefan Grosskinsky}

\affiliation{Centre for Complexity Science, University of Warwick, Coventry CV4 7AL, UK}

\author{Rosemary~J. Harris}

\affiliation{School of Mathematical Sciences, Queen Mary University of London, London E1 4NS, UK}

% Formatting and style data

\begin{abstract}
Information thermodynamics provides a framework for studying the effect of feedback loops on entropy production. It has enabled the understanding of novel thermodynamic systems such as the information engine which can be seen as a modern version of `Maxwell's D\ae mon', whereby the feedback controller is acting as a D\ae mon, processing information gained about the system in order to do work. Here, we analyse a simple model of such an engine and provide a detailed analysis of its fluctuation properties, including the large deviations of information. We find an exact expression of the large deviation rate function for a two-site version of our model, and provide an approximate analysis for larger systems which is corroborated by simulation data.
\end{abstract}

% Title page. This contains a title/author/date section, plus the abstract.

\maketitle

\newpage

% Main report from here on.

%%%%%%%%%%%%%%%%%%%%%%%%%%%%%%%%%%%%%%%%%%%%%%
%%%%%%%%%%%%%%%%%%%%%%%%%%%%%%%%%%%%%%%%%%%%%%
\section{Introduction}\label{Sec:Intro}

The groundwork for information thermodynamics was laid down by Maxwell as part of his now infamous thought experiment `Maxwell's D\ae mon'~\cite{thomson1879sorting}. In the experiment, a sentient agent monitors the motion of thermal particles inside a partitioned container. By operating a small gate in the partition, fast-moving particles are allowed to move to one side of the partition while slow moving particles are allowed to move to the other side, thus heating up the first side and decreasing the overall entropy of the system. 

While first conceived to elucidate the statistical subtleties of the 2\textsuperscript{nd} law of thermodynamics, the experiment has since sparked many debates~\cite{szilard1929entropieverminderung,brillouin1951maxwell,landauer1961irreversibility,bennett1982thermodynamics,leff2010maxwell} on the nature of the perceived violation of the 2\textsuperscript{nd} law. Note that these violations relate to the ensemble average and are different from the temporary violations occuring on the level of individual trajectories or events~\cite{gallavotti1995dynamical,evans1993probability,evans1994equilibrium}.

The culmination of nearly 150 years of discussions about Maxwell's D\ae mon has been the framework of information thermodynamics~\cite{sagawa2010generalized}, which illuminates the profound importance of information in thermodynamics~\cite{seifert2012stochastic,abreu2012thermodynamics}. By including the effect of system memory~\cite{zhou2010minimal} and/or information processing (often thought to be performed by the feedback device, referred to as a `D\ae mon'~\cite{maruyama2009colloquium}), the 2\textsuperscript{nd} law can be reformulated to include information entropy and used to study the operation of finite-time thermodynamic systems~\cite{berrythermodynamic} such as information heat engines~\cite{jayannavar1996simple}.

`Information heat engines' are a class of thermodynamic systems that use information processing to do thermodynamic work without the need for a change in free energy~\cite{toyabe2010experimental,park2013heat}. Methods such as feedback control allow the engine to use information gained about a physical system to decrease the system entropy and hence extract useful work from the system~\cite{lloyd1997quantum}. This does not constitute a violation of the 2\textsuperscript{nd} law as it is understood that the operation of the feedback device entails an amount of entropy production at least equal and opposite to that change in the system~\cite{piechocinska2000information,landauer1961irreversibility}. 

A quantitative relationship between entropy and information is provided by the information thermodynamic framework, which gives the universal upper bound on the mean negative entropy production that can be obtained by feedback control~\cite{sagawa2008second}. 

To be precise, the `2\textsuperscript{nd} law of information thermodynamics' states that the entropy production $S_t$, of a system up to time $t$ is related to the information, $I_t$, gained by the D\ae mon via the inequality~\cite{touchette2000information}
\begin{equation}
\langle S_t \rangle \geq - \langle I_t\rangle,
\label{Eq:2ndlaw}
\end{equation}
where the angle brackets denote the ensemble average. In fact, this turns out to be a corollary of the generalised integral fluctuation theorem~\cite{sagawa2010generalized} (itself a generalisation of the Jarzynski equality~\cite{jarzynski1997nonequilibrium}),
\begin{equation}
\langle e^{-S_t-I_t} \rangle = 1.
\label{Eq:GIFR} 
\end{equation}
this implies that for $I_t \neq 0$,
\begin{equation}
\langle e^{-S_t} \rangle \neq 1,
\label{Eq:SIFR} 
\end{equation}
that is, the standard integral fluctuation theorem does not hold in the presence of feedback~\cite{sagawa2012nonequilibrium}. These results have been experimentally verified in small systems, where thermal fluctuations have a strong influence~\cite{toyabe2010experimental,berut2012experimental,koski2014experimental}. In previous theoretical studies, the quantity $I_t$ and its relation to $S_t$ has been discussed in the context of Langevin equations and continuous-time Markov chains~\cite{horowitz2014thermodynamics}, and the mutual information between the feedback controller and the stochastic system has been considered for systems with discrete events~\cite{horowitz2013imitating}. 

Here, we study an abstract model of an `information motor'~\cite{cao2004feedback,lopez2008realization,suzuki2009one,horowitz2013imitating}. The information motor discussed here is type of ratchet that is able to move a single particle against a bias using only the particle's own random motion and a feedback mechanism. Specifically, this model allows us to demonstrate a method for calculating the information $I_t$ in a discrete-time feedback system and to study its fluctuation properties.

The paper is structured as follows. Sec.~\ref{Sec:Engines} contains an overview of the existing information thermodynamic framework. In this section we also detail the method used for calculating the information gained (in a single measurement). In Sec.~\ref{Sec:Model} we describe our simple model of a Maxwell's D\ae mon type feedback system. In Sec.~\ref{Sec:LargeDeviations} we discuss the fluctuations of information on the level of individual trajectories and how to obtain large deviation rate functions. In Sec.~\ref{Sec:2SiteModel} we obtain exact expressions for the large deviation rate function in a two-site version of our model. In Sec.~\ref{Sec:LStepModel} we then give a detailed approximate analysis and obtain numerical results for larger information engines. In Sec.~\ref{Sec:Discussion} we conclude by summarising our results and discussing potential further work and open questions. The appendices contain further details of the two-site system and properties of individual trajectories.
%%%%%%%%%%%%%%%%%%%%%%%%%%%%%%%%%%%%%%%%%%%%%%
%%%%%%%%%%%%%%%%%%%%%%%%%%%%%%%%%%%%%%%%%%%%%%
\section{Framework}\label{Sec:Engines}

For simplicity's sake, let us consider a stochastic system evolving in discrete time and having states in some finite state space of size $L$. The state of the system at a time $s$ is represented as a random variable $X_s$ with a specific realisation denoted by $x_s$. A trajectory of the system is written $\mathbf{X}_t= \{X_s\}_{s=0}^t$ with a specific realisation denoted by $\mathbf{x}_t$. The probability of a transition between states $x$ and $x'$ is written as $\omega(x \to x')$. 

For a system subject to general feedback, we consider that the transition probabilities are determined by some other parameter referred to as the control parameter. In purely `open-loop' control, the control parameter is independent of the system state. However, in the case of `closed loop' or `feedback' control, the system's evolution influences the control parameter in a closed causal loop~\cite{astrom2010feedback}.

In the case of Maxwell's D\ae mon, the D\ae mon is identified as a feedback controller~\cite{touchette2000information} whose activity can be described by two processes, measurement and control. The measurement is the process that allows the controller to select a control parameter to `feed' back into the system via the control process as described above. The measurement is represented in a similar fashion to the system trajectory, and is written as $\mathbf{Y}_t=\{Y_s\}_{s=0}^t$. The measurement $Y_s$ at time $s$ only depends on the current state $X_s$ and so we denote,
\begin{equation}
p(y_s\given x_s) \coloneqq \mathbb{P} \left[Y_s=y_s\given X_s=x_s\right]
\label{Eq:Error} 
\end{equation}
as the probability of obtaining outcome $y_s$ given that the system is in state $x_s$. Here we assume an injective mapping between measurement outcomes and control parameters, that is, a given measurement $y_s$ determines a unique control parameter and thus along with the departure state $x_s$ determines the probability of transitions to the next state $x_{s+1}$; we write this as
\begin{equation}
\omega(x_s \to x_{s+1}\given y_s).
\end{equation} 

The conditional distribution in~\eqref{Eq:Error} is derived from $\mathbb{P} $, the path space measure of the full process $\{(\mathbf{X}_s,\mathbf{Y}_s)\}_{s=0}^t$. This process is a Markov chain on the state space given by $X$ and $Y$ pairs and can be described by the transition matrix 
\begin{equation}
\Omega((x,y)\to(x',y')) \coloneqq \omega(x\to x'\given y)p(y'\given x').
\label{Eq:MarkovMatrixDef}
\end{equation}
We also write $P(\mathbf{x}_s)=\mathbb{P}\left[\mathbf{X}_s=\mathbf{x}_s\right]$ and $P(\mathbf{y}_s)=\mathbb{P}\left[\mathbf{Y}_s=\mathbf{y}_s\right]$ for the marginal distributions of the process and measurement trajectories respectively. Note that, while the measurements $Y_s$ are conditionally independent given the path $\mathbf{X}_t$, the marginal measurement process $\mathbf{Y}_t$ exhibits correlations after integrating out $\mathbf{X}_t$ and is not a sequence of independent identically distributed (i.i.d.) random variables. Extending the definition in~\cite{crooks1999entropy}, the entropy production at time $s$ as a function of a given trajectory $\left(\mathbf{x}_t,\mathbf{y}_t\right)$ in a feedback system is
\begin{equation}
\Delta S_s = \ln{\frac{\omega(x_s \to x_{s+1} \given y_s)}{\omega(x_{s+1} \to x_{s} \given y_s)}}.
\label{Eq:DeltaSDef}
\end{equation}

As mentioned above, for a system with feedback it is also necessary to quantify and study the information gained through measurement. When considering the information gained in a single measurement, we follow~\cite{horowitz2010nonequilibrium} and use the `change in uncertainty', an information theoretic quantity that quantifies the information gained upon making an observation of some process. For a single measurement this is given by
\begin{align}
\Delta I_s &= \ln{\frac{p(y_s \given x_s)}{P(y_s \given\mathbf{y}_{s-1})}}, \nonumber\\
&= \ln{\frac{p(y_s\given x_s)P(\mathbf{y}_{s-1})}{P(\mathbf{y}_s)}},
\label{Eq:DeltaIDef}
\end{align}
for all $s\geq1$, where 
\begin{equation}
P(y_s \given \mathbf{y}_{s-1}) = \mathbb{P} \left[Y_s{=}y_s \given Y_0{=}y_0,\ldots,Y_{s-1}{=}y_{s-1}\right].
\label{Eq:YsGivenY}
\end{equation}

Defining $\pi_0(x_0) = \mathbb{P} \left[X_0 = x_0\right]$, the denominator term in~\eqref{Eq:DeltaIDef} is given by
\begin{align}
P(\mathbf{y}_s) &= \sum_{\mathbf{x}_{s+1}} \mathbb{P}\left[ \mathbf{X}_{s+1} =\mathbf{x}_{s+1} ,\mathbf{Y}_s =\mathbf{y}_s\right], \nonumber\\
&= \sum_{\mathbf{x}_{s+1}} \pi_0(x_0)\prod_{u=0}^{s}\omega(x_u\to x_{u+1} \given y_u)p(y_u\given x_u).
\label{Eq:PYDef}
\end{align}
We can evaluate~\eqref{Eq:PYDef} by writing the sum as a matrix product, representing the terms in the sum as the elements of matrices
\begin{equation}
(M_{y_s})_{x,x'} \coloneqq \omega(x \to x' \given y_s)p(y_s \given x).
\label{Eq:MMatrixDef} 
\end{equation}
We write the initial probability distribution as a vector $\bra{\pi_0} = (\pi_0(1),\pi_0(2),\ldots,\pi_0(L))$ and also define a summation vector of length $L$, $\ket{1} = (1,1,\ldots,1)^T$. We can then write $P(\mathbf{y}_t)$ as
\begin{equation}
P(\mathbf{y}_t) = \bra{\pi_0}\prod_{s=0}^t M_{y_s} \ket{1}.
\label{Eq:PYwithM}
\end{equation}

For any given series of measurements $\mathbf{y}_t$ we can calculate $P(\mathbf{y}_t)$. Each of the matrices has a size $L\times L$ and we must multiply $t$ of these matrices to calculate the information obtained up to time-step $t$. Eq.~\eqref{Eq:DeltaIDef} can then be written as
\begin{equation}
\Delta I_s = \ln{p(y_s\given x_s)\frac{\bra{\pi_0}\prod_{s'=0}^{s-1}M_{y_s'}\ket{1}}{\bra{\pi_0}\prod_{s'=0}^{s}M_{y_s'}\ket{1}}}.
\label{Eq:DeltaIwithM}
\end{equation}
We will see in Sec.~\ref{Sec:2SiteModel} that this representation of $\Delta I_s$ also clarifies potential cancellation of terms in the products.

Here we are interested in the fluctuations of the total information gained $I_t$, which is obtained by summing terms~\eqref{Eq:DeltaIDef} along a given trajectory $(\mathbf{x}_t,\mathbf{y}_t)$. This yields
\begin{align}
I_t(\mathbf{x}_t,\mathbf{y}_t) &= \sum_{s=0}^t \ln{\frac{p(y_s \given x_s)}{P(y_s \given \mathbf{y}_{s-1})}}, \nonumber\\
&= \ln{\frac{P(\mathbf{y}_t\given \mathbf{x}_t)}{P(\mathbf{y}_t)}}.
\label{Eq:ItDef}
\end{align}
Here we have used the conditional independence of the measurements $P(\mathbf{y}_t\given\mathbf{x}_t) {=} \prod_{s=0}^t p(y_s\given x_s)$, defined $\mathbf{y}_{-1}$ such that $P(y_0 \given \mathbf{y}_{-1}) \equiv P(y_0)$ and used $P(\mathbf{y}_t) = \prod_{s=0}^t P(y_s\given \mathbf{y}_{s-1})$.

%%%%%%%%%%%%%%%%%%%%%%%%%%%%%%%%%%%%%%%%%%%%%%
%%%%%%%%%%%%%%%%%%%%%%%%%%%%%%%%%%%%%%%%%%%%%%
\section{Model}\label{Sec:Model}

We consider a model that abstractly resembles a recent experimental set-up involving a colloidal particle rotating in an electric field~\cite{toyabe2010experimental}. The experimental system demonstrated a type of particle ratchet where a field can be switched and shifted along with the motion of the particle in order to block it from moving. By ratcheting in this way, the particle's own thermal motion can be used to do work.

Our model is comprised of a random walk on a one dimensional lattice with a movable barrier. The random walk acts as an analogy to the colloidal particle in the experiment, with the walker's random motion modeling the thermal motion of the experimental particle. We model in discrete space, as in the real experiment there was a single coarse-grained measurement, essentially allowing the identification of discrete `states'.

Furthermore, the measurements were performed at regular intervals which allows the whole feedback process to be thought of in discrete time-steps. In our model, the random walker moves between sites on a lattice of size $L$ with periodic boundary conditions and probabilities $q$ and $p$ of jumping respectively left and right at each time-step, such that $p+q=1$. The random walker's motion is then described by a single parameter $p$, which in the case where $p\neq q$ describes a system with a bias in one direction. Without loss of generality, we consider $q>p$ for biased systems. We label jumps left as `down', and jumps right as `up' as though the particle were moving in a potential.

In the spirit of the feedback process described in~\cite{toyabe2010experimental}, at each time-step a measurement is made of the particle position and the barrier is moved to the measured location of the particle in an attempt to prevent it from moving down. When the particle attempts to jump `through' the barrier, it instead remains at the same site. If the measurement is always correct, the particle can only ever jump up and so the system acts as a perfect ratchet. However, if the measurement is incorrect then the barrier will be placed incorrectly and will either not affect the particle's motion or will act as a blockade for jumps up. 
 \begin{figure}
\captionsetup{justification=raggedright, singlelinecheck=false}
\includegraphics[width=0.45\textwidth]{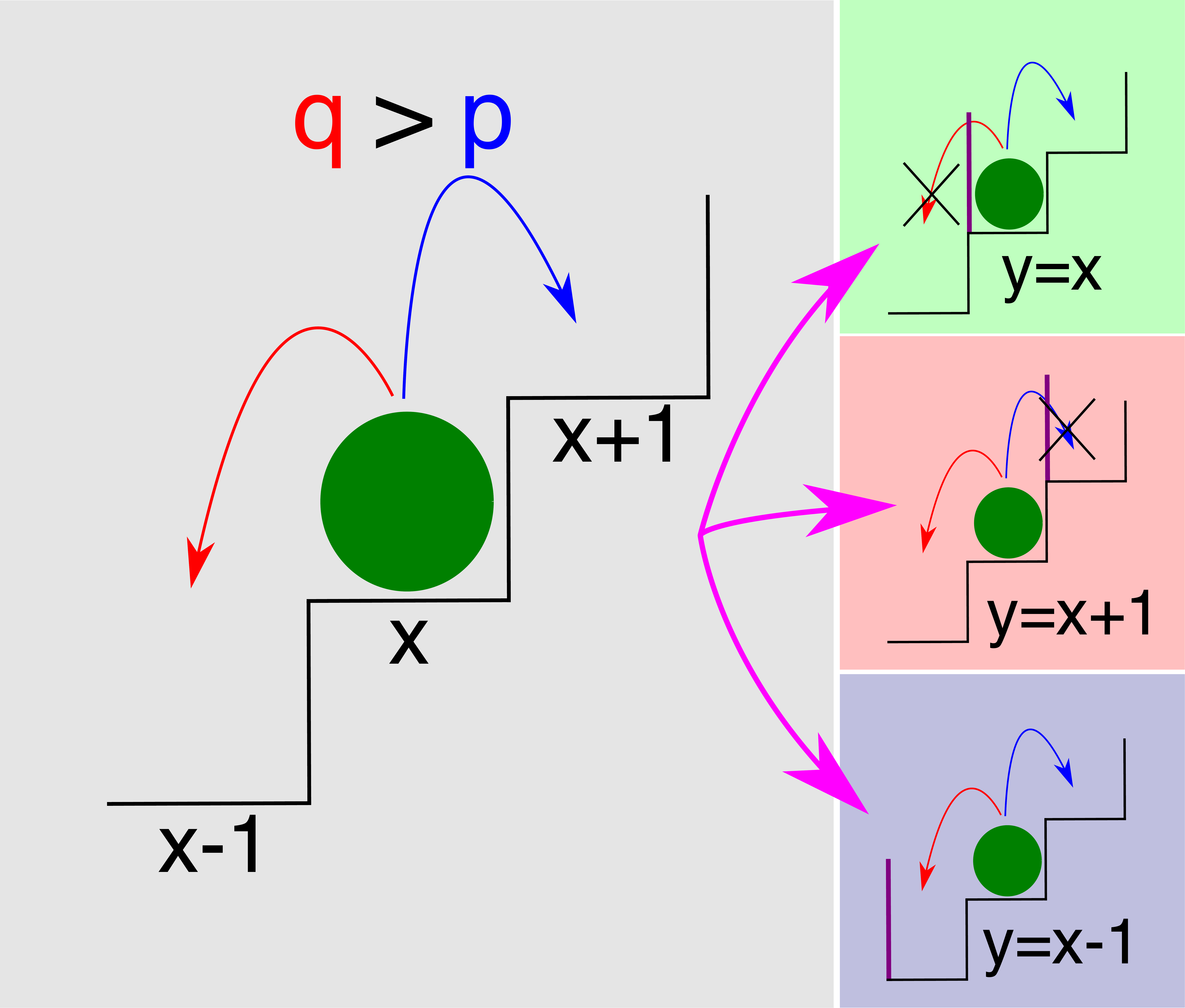}
\caption{Schematic of model. The feedback mechanism makes a measurement of the particle position and places a barrier according to the outcome of the measurement. If $y=x$ or $y=x+1$ then the barrier influences the particle movement by preventing certain jumps as in the top and middle right. For all other measurement values, the barrier has no effect and the particle moves freely as in the bottom right.}
\label{Fig:Daemon}
\end{figure}
The system is initialised by first choosing a site $x_0$ uniformly from the lattice sites and then performing the measurement process to obtain a $y_0$. Fig.~\ref{Fig:Daemon} is a schematic diagram of the system where the bias is represented by showing the lattice as a staircase. The three possible results of the action of the feedback device are shown, including the situation where the feedback has no effect on the particle motion.

The model can be described by three parameters: the lattice size $L$, the motion bias $p$ and the measurement accuracy $r$ which we define as
\begin{equation}
r\coloneqq p(x\given x) \quad \forall ~ x\in\{ 1,\ldots ,L\}.
\end{equation}
The corresponding error probabilities
\begin{equation}
w\coloneqq p(y\given x) \quad \forall ~ y, x\in\{ 1,\ldots ,L\},~y\neq x,
\end{equation}
are related to $r$ via the normalisation condition $w = (1-r)/(L-1)$. The measurement error is then independent of which site $x$ the walker is at and all incorrect measurements (i.e.\ any $y\neq x$) are equally likely.  

We consider the case of `accurate' measurements ($r > w$) as in this regime the measurements can be used to make useful inferences about the system state and the information gained through measurement can be used in the operation of the information engine. Contrastingly, in the special case $r=w=1/L$, the joint probability in~\eqref{Eq:PYDef} factorises and~\eqref{Eq:ItDef} is always zero; no information is ever gained by the D\ae mon and hence the system reduces to a `lazy' random walk. When $r<w$ it is possible for the D\ae mon to gain information, but we do not study this regime as it is not clear in general how to utilise the information gained from a measurement device that measures wrongly more frequently than correctly.

These three parameters, $p$, $r$ and $L$ completely characterise the model. At each time-step the contribution to the entropy production is given by~\eqref{Eq:DeltaSDef} and is $\Delta S_s \in \{-\ln{\left(p/q\right)},0,\ln{\left(p/q\right)}\}$, where the non-zero terms correspond to successful jumps down or up and $0$ is the entropy produced if the particle attempts to move through the barrier. The information gained by the feedback controller is given by~\eqref{Eq:DeltaIwithM}. All three parameters determine the average particle current
\begin{equation}
\frac{\langle J_t \rangle}{t} = pr+pw(L-2)+qw(1-L),
\end{equation}
where positive current is in the rightward direction. Our choice of parameters ($q\geq p$ and $r>w$) can produce a positive average current (current against the bias), when in the absence of feedback, zero or negative current would be expected.

\begin{figure}
\captionsetup{justification=raggedright, singlelinecheck=false}
\includegraphics[width=0.45\textwidth]{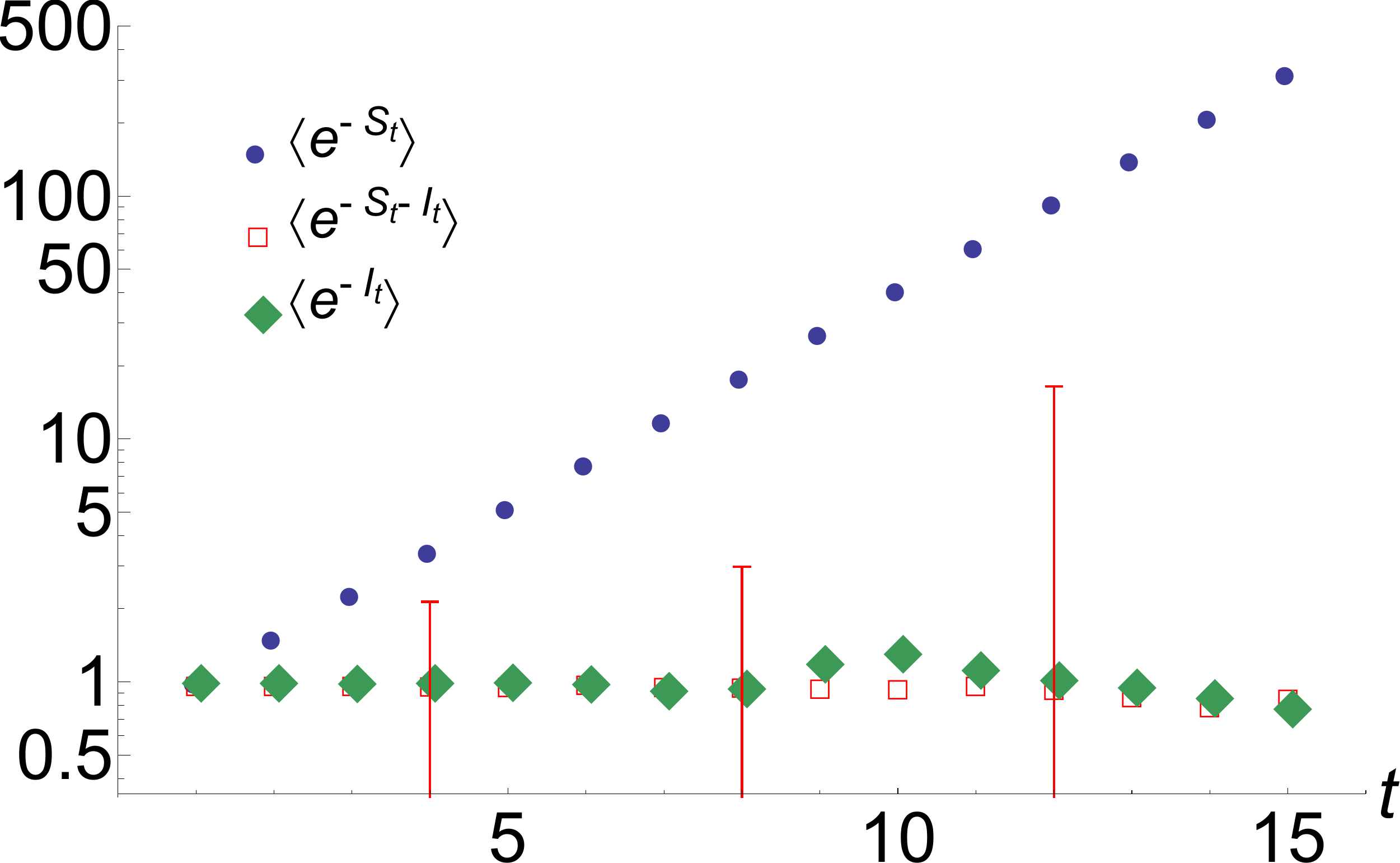}
\caption{Numerical tests of~\eqref{Eq:GIFR},~\eqref{Eq:SIFR} and~\eqref{Eq:IFRI} in semi log scale. Parameter values are $L=3$, $p=0.2$, $r=0.9$. Averaged over $10^7$ realisations. Error bars indicate the standard error of mean of $\langle e^{-S_t-I_t}\rangle$ at representative points.}
\label{Fig:IFRCheck}
\end{figure}

As a preliminary to Sec.~\ref{Sec:LargeDeviations}, we numerically check the generalised integral fluctuation theorem~\eqref{Eq:GIFR}, show that the standard integral fluctuation theorem~\eqref{Eq:SIFR} does not hold, and check the equality,
\begin{equation}
\langle e^{-I_t}\rangle = 1,
\label{Eq:IFRI}
\end{equation} 
which follows from the definition~\eqref{Eq:ItDef}. Fig.~\ref{Fig:IFRCheck} confirms these (in)equalities at different times $t$ with numerical data obtained from Monte Carlo simulation of a biased three-site system. Since the means of exponential quantities are determined by rare events, fluctuations are large as indicated by the error bars on representative points.

%%%%%%%%%%%%%%%%%%%%%%%%%%%%%%%%%%%%%%%%%%%%%%
%%%%%%%%%%%%%%%%%%%%%%%%%%%%%%%%%%%%%%%%%%%%%%
\section{Information fluctuations and large deviations}\label{Sec:LargeDeviations}

\subsection{Fluctuations of $\Delta I_s$}\label{Sec:DeltaIsFlucs}
Having numerically checked the ~\eqref{Eq:GIFR},~\eqref{Eq:SIFR} and~\eqref{Eq:IFRI}, we now proceed to investigate in more detail the fluctuation properties of the information gained by the D\ae mon. The variables $I_t / t$ and $\Delta I_s$ are of interest, being the time-averaged information gain and instantaneous change in uncertainty, respectively. On the level of an individual trajectory there are three types of events possible, given the system dynamics, that influence $\Delta I_s$. These are roughly described as follows,
\begin{itemize}
\item Correct measurements
\item Incorrect measurements that are recognisable as such
\item Incorrect measurements that are not recognisable as such
\end{itemize}
We observe from the analysis of trajectories in App.~\ref{App:TrajectoryAnalysis} that correct measurements yield positive amounts of information. Correctly observed jumps against the bias yield more information than blocked jumps in the direction of the bias. For symmetric systems the difference in $\Delta I_s$ between jumps left and right is small. Series of correct measurements yield one of two baseline values for information gain $\Delta I_s$ that correspond to jumps up and blocked jumps down (see App.~\ref{App:TrajectoryAnalysis}).

If an incorrect measurement is made, then the ratio $P(\mathbf{y}_{s-1})/P(\mathbf{y}_s)$ in Eq.~\eqref{Eq:DeltaIDef} changes and the amount of information gained is zero if the current measurement is not compatible with the previous measurement (e.g.\ the particle looks like it has jumped `through' the barrier or jumped more than one site). A sequence of incorrect measurements can lead to negative values in information gain, interpreted as a change towards greater uncertainty of the system state.

Whenever a series of incorrect measurements is made, the next series of correct measurements gains large positive amounts of information that partially retrieve the `lost' information of the incorrect measurements. As these correct measurements are made, the ratio $P(\mathbf{y}_{s-1})/P(\mathbf{y}_s)$ relaxes and $\Delta I_s$ returns to a baseline value which we explain in detail in App.~\ref{App:TrajectoryAnalysis}. The information theoretic interpretation is that successive correct measurements allow the observer to infer which measurement was incorrect. However, for consistent incorrect measurements which are not detectable the information lost cannot be retrieved. A discussion of this with an example is given in App.~\ref{App:TrajectoryAnalysis}.

Large positive deviations of $I_t / t$ are not generated by an accumulation of the largest values of $\Delta I_s$ as these are necessarily preceded by large negative values as described above. They are instead generated by strings of correct measurements which each generate less information than the largest values of $\Delta I_s$. 

In contrast, large negative deviations are generated by sequences of incorrect measurements which happen to represent a possible system trajectory. In this case large negative values of $\Delta I_s$ can accumulate and are not compensated by subsequent large positive values. In general, since the baseline values discussed above depend on the trajectory in case of correct measurements, atypical trajectories $\mathbf{x}_t$ also play a role in the realization of large deviations of $I_t /t$.

\subsection{Large Deviation Analysis}\label{Sec:LargeDevAnalysis}
In order to study the fluctuation properties of $I_t / t$, we follow standard methods (see, e.g.\ \cite{touchette2009large}) and start by assuming that $I_t / t$ obeys a large deviation principle of the form,
\begin{equation}
\mathbb{P}\left[I_t \approx it\right] \sim e^{-E(i)t},
\label{Eq:LargeDevDef}
\end{equation}
as $t$ approaches infinity, where $E(i)$ is the `large deviation rate function'~\cite{varadhan1966asymptotic}. The rate function tells us about the fluctuation properties of the variable $I_t / t$, and allows us to quantify how exponentially unlikely a given fluctuation away from the mean is, in the long-time limit. 

In order to calculate the rate function, we first consider the scaled cumulant generating function (SCGF),
\begin{equation}
\xi(k) \coloneqq \lim_{t\to\infty}-\frac{1}{t}\ln{G(k)},
\label{Eq:SCGFDef}
\end{equation}
where $G(k)$ is the moment generating function, 

\begin{equation}
G(k) \coloneqq \langle e^{-k I_t} \rangle = \int_{-\infty}^{\infty}e^{-k u}\mathbb{P}\left[I_t= u\right]\mathrm{d}u.
\label{Eq:GenFn}
\end{equation}
The rate function is then the Legendre-Fenchel transform of the SCGF
\begin{equation}
E(i) = \operatorname*{sup}_{k\in\mathbb{R}}\left[\xi(k)- k i\right].
\label{Eq:LegendreFenchelTransform}
\end{equation}

By rewriting an expectation of the form~\eqref{Eq:GenFn}, the SCGF can often be obtained as the logarithm of the principal eigenvalue of the Markov transition matrix which is weighted to count the relevant quantity~\cite{schuster2013nonequilibrium3}. This approach can be applied if the measured quantity depends additively on transitions along a system trajectory, as is the case with particle current. Equation~\eqref{Eq:DeltaIDef} shows that this is in general not the case for information, since the gain in a given measurement depends on the entire measurement history up until that point. Hence transitions on the enlarged state space of pairs $(x_s, y_s)$ cannot be associated with specific values of $\Delta I_s$. 

However, in the next section we show that for $L=2$, $\Delta I_s$ can be simplified using Eq.~\eqref{Eq:DeltaIwithM} to an expression that only depends on the departure and target states in a single transition, and the above approach can be applied by weighting the transition matrix~\eqref{Eq:MarkovMatrixDef} to count information gain leading to an exact analytical expression of the large deviation rate function. This simplification does not hold for $L\geq 3$, and in Sec.~\ref{Sec:LStepModel} we describe approximate methods for obtaining the rate function by formulating a one-step Markov model for the sequence of $\Delta I_s$.

%%%%%%%%%%%%%%%%%%%%%%%%%%%%%%%%%%%%%%%%%%%%%%
%%%%%%%%%%%%%%%%%%%%%%%%%%%%%%%%%%%%%%%%%%%%%%
\section{Exact computation for two-site system}\label{Sec:2SiteModel}

For a system with two sites labeled $1$ and $2$, the matrices used in~\eqref{Eq:PYwithM} are
\begin{equation}
M_{1} = \left(
\begin{array}{cc}
 q r & p r \\
 q w & p w  \\
\end{array}
\right)
\end{equation}
\begin{equation}
M_{2} = \left(
\begin{array}{cc}
 p w & q w \\
 p r & q r  \\
\end{array}
\right)
\end{equation}
corresponding to the two measurement outcomes. For this system, the $M_y$ matrices for any $L$ are similarity transforms of one another and have the same spectrum. For the $L=2$ case these matrices have only one non-zero eigenvalue $\lambda=qr+pw$, allowing them both to be written as tensor products on the eigenspace of the corresponding eigenvector. That is, we can write $M_{y} = \ket{v_y^r}\bra{v_y^l}$ where $\bra{v_y^l}$ and $\ket{v_y^r}$ are the left and right eigenvectors of $M_y$ with respect to $\lambda$. The matrices can be rewritten as
\begin{equation}
M_{1} = \ket{v_1^r}\bra{v_1^l} = \left(
\begin{array}{c}
 r \\
w \\
\end{array}
\right)
\left(
\begin{array}{cc}
 q & p \\
\end{array}
\right)
\end{equation}
and
\begin{equation}
M_{2} = \ket{v_2^r}\bra{v_2^l} = \left(
\begin{array}{c}
w \\
r \\
\end{array}
\right)
\left(
\begin{array}{cc}
 p & q \\
\end{array}
\right).
\end{equation}

Writing the matrices in this way shows that all but the final term of the upper product in~\eqref{Eq:DeltaIwithM} cancel with terms in the lower product, leaving an inner product between two vectors:
\begin{align}
\frac{P(\mathbf{y}_s)}{P(\mathbf{y}_{s-1})} &= \frac{\bra{\pi_0}M_{y_{0}}M_{y_1}\ldots M_{y_{s-1}}M_{y_s}\ket{1}}{\bra{\pi_0}M_{y_{0}}M_{y_1}\ldots M_{y_{s-1}}\ket{1}} \nonumber\\
&=\frac{\bra{\pi_0}\ldots\ket{v_{y_{s-1}}^r}\braket{v_{y_{s-1}}^l \given v_{y_s}^r}\braket{v_{y_s}^l \given 1}}{\bra{\pi_0}...\ket{v_{y_{s-1}}^r}\braket{v_{y_{s-1}}^l \given 1}} \nonumber\\
&= \braket{v_{y_{s-1}}^l \given v_{y_s}^r}.
\label{Eq:PYcancellation}
\end{align}
$P(\mathbf{y}_s)/P(\mathbf{y}_{s-1})$ then can take four values corresponding to the values taken by $y_{s-1}$ and $y_s$, i.e., $y_{s-1},y_s\in \{1,2\}$. However, as the two $M_y$ matrices are permutations of one another, only two distinct values can be obtained, when $y_{s}\neq y_{s-1}$ or when $y_s = y_{s-1}$. The change in information upon making a measurement in this two-site system is
\begin{equation}
\Delta I_s = \ln{\frac{p(y_s \given x_s)}{\braket{v_{y_{s-1}}^l \given v_{y_s}^r}}},
\label{Eq:2SiteDeltaIs}
\end{equation}
which for a given $x_s$ depends only on the previous and current measurements $y_{s-1}$ and $y_s$. From~\eqref{Eq:2SiteDeltaIs} it can further be deduced that the process $\{\Delta I_s\}_{s=0}^t$ is simply a sequence of i.i.d.\ random variables. This simplification only holds for $L=2$ and is demonstrated in App.~\ref{App:IID}. 

As $p(y_s\given x_s)$ has two possible values and $\braket{v_{y_{s-1}}^l \given v_{y_s}^r}$ also has two possible values, $\Delta I_s$ takes four possible values. Specifically, these are
\begin{eqnarray}
\ln{\frac{r}{pr+qw}} \coloneqq a, ~ \label{Eq:2SiteDeltaIsA}\ln{\frac{r}{qr+pw}} \coloneqq b,
\label{Eq:2SiteDeltaIsB}
\end{eqnarray}
for correct measurements made after jumps in the up and down directions (whether blocked or not), respectively. For the same cases followed by incorrect measurements, $\Delta I_s$ takes the values
\begin{eqnarray}
\ln{\frac{w}{pr+qw}} \coloneqq c, ~ \label{Eq:2SiteDeltaIsC}\ln{\frac{w}{qr+pw}} \coloneqq d.
\label{Eq:2SiteDeltaIsD}
\end{eqnarray}
\begin{figure}
\captionsetup{justification=raggedright, singlelinecheck=false}
\includegraphics[width=0.45\textwidth]{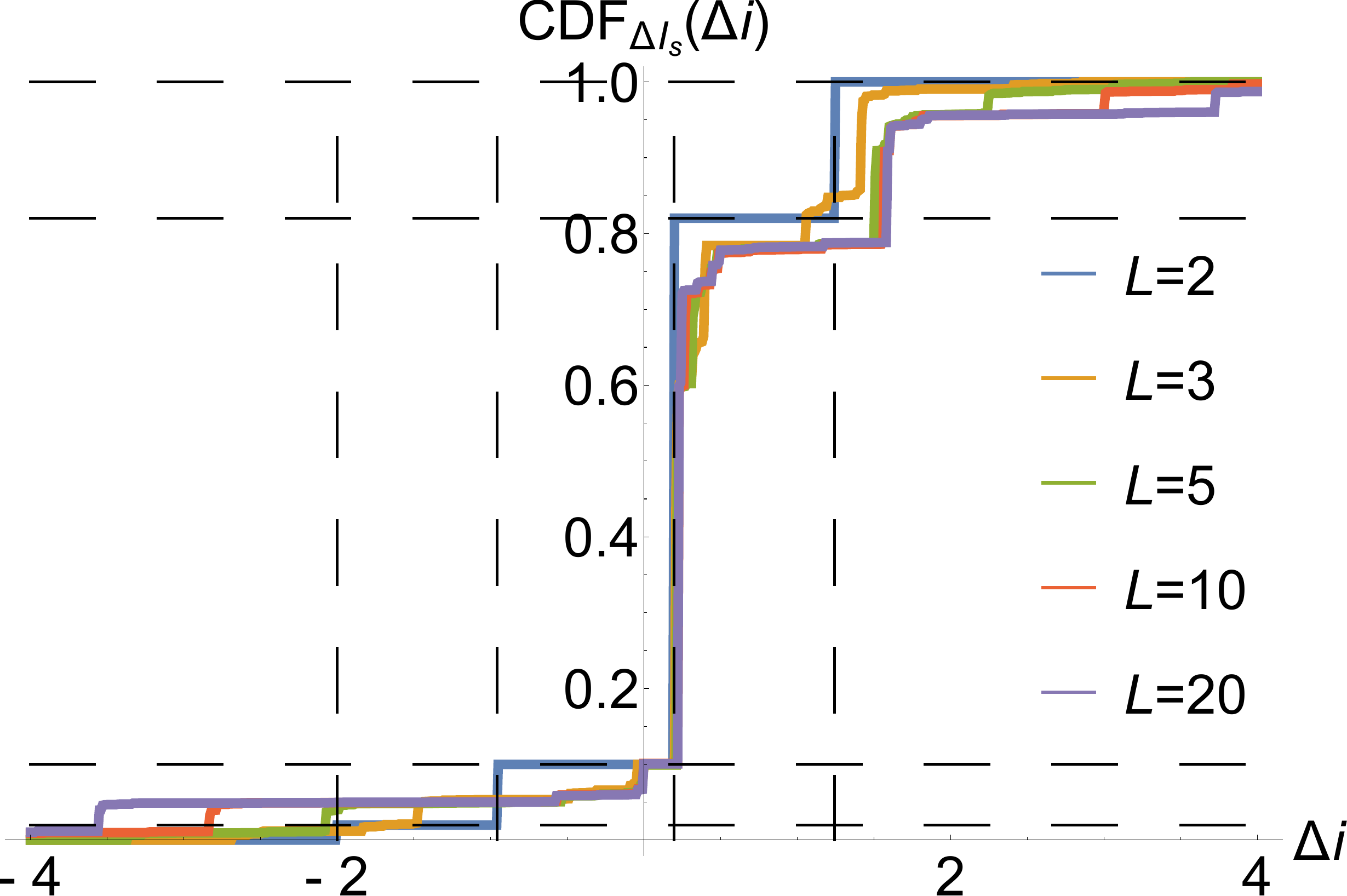}
\caption{Cumulative density function for $\Delta I_s$ for different system sizes $L$ with $p=0.2$ and $r=0.9$. Dashed lines give theoretical values for $L=2$ as given by~\eqref{Eq:2SiteDeltaIsA} and\eqref{Eq:2SiteDeltaIsD}.}
\label{Fig:CDFDeltaIs}
\end{figure}

The values taken by $\Delta I_s$ can be associated with transitions on the state space $\left(x_s,y_s\right)$, which is described by the transition matrix~\eqref{Eq:MarkovMatrixDef}. For $L=2$ this matrix is 
\begin{equation}
\Omega = 
\left(
\begin{array}{cccc}
 q r & q w & p w & p r \\
 p r & p w & q w & q r \\
 q r & q w & p w & p r \\
 p r & p w & q w & q r \\
\end{array}
\right).
\label{Eq:2SiteMarkovMatrix}
\end{equation} 
The values in~\eqref{Eq:2SiteDeltaIsA} and \eqref{Eq:2SiteDeltaIsD}, along with the probabilities of these events occurring, given by the transition matrix~\eqref{Eq:2SiteMarkovMatrix}, are enough to determine the cumulative density function (CDF) of the random variable $\Delta I_s$ for all $s\geq1$
\begin{equation}
\operatorname{CDF}_{\Delta I} (\Delta i)\coloneqq\mathbb{P} \left[\Delta I \leq \Delta i\right] \quad \forall ~ \Delta i\in\mathbb{R},
\label{Eq:CDFDef}
\end{equation}
which is plotted in Fig.~\ref{Fig:CDFDeltaIs} and compared with numerical data.

\begin{figure}
\captionsetup{justification=raggedright, singlelinecheck=false}
\includegraphics[width=0.45\textwidth]{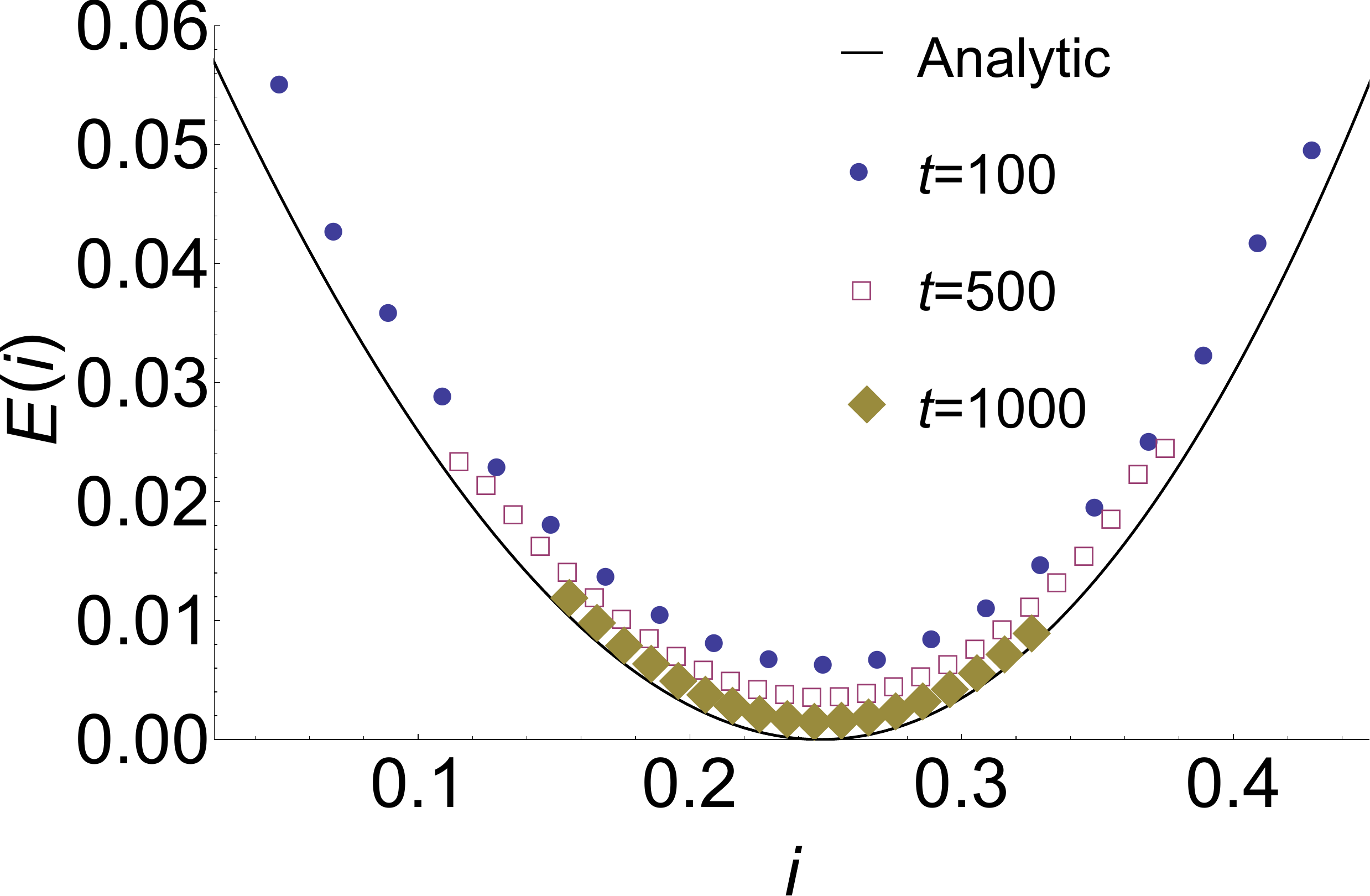}
\caption{Large deviation rate function~\eqref{Eq:LegendreFenchelTransform} for $L=2$, $p=0.2$, and $r=0.9$. The solid black line shows the theoretical curve obtained from Eq.~\eqref{Eq:2SiteSCGF} and~\eqref{Eq:LegendreFenchelTransform}. Points represent data sampled at different finishing times $t$.}
\label{Fig:2SiteRateFn}
\end{figure}

It is possible to weight the Markov transition matrix~\eqref{Eq:2SiteMarkovMatrix} with the values of $\Delta I_s$ from~\eqref{Eq:2SiteDeltaIsA} and \eqref{Eq:2SiteDeltaIsD} to obtain the tilted matrix,
\begin{equation}
\Omega'(k) = 
\left(
\begin{array}{cccc}
 e^{-b k} q r & e^{-c k} q w & e^{-d k} p w & e^{-a k} p r \\
 e^{-a k} p r & e^{-d k} p w & e^{-c k} q w & e^{-b k} q r \\
 e^{-b k} q r & e^{-c k} q w & e^{-d k} p w & e^{-a k} p r \\
 e^{-a k} p r & e^{-d k} p w & e^{-c k} q w & e^{-b k} q r \\
\end{array}
\right),
\end{equation}
which has principal eigenvalue
\begin{align}
\lambda(k) =& e^{-a k} p r+e^{-b k} q r +e^{-c k} q w+e^{-d k} p w.
\label{Eq:2SiteSCGF}
\end{align}
The logarithm of~\eqref{Eq:2SiteSCGF} is taken as the SCGF $\xi(k)$ and Legendre transformed according to~\eqref{Eq:LegendreFenchelTransform} into the rate function $E(i)$. Fig.~\ref{Fig:2SiteRateFn} shows this rate function plotted with data from simulation for a two-site system.  In the long-time limit the data converge well to the rate function.
%

%%%%%%%%%%%%%%%%%%%%%%%%%%%%%%%%%%%%%%%%%%%%%%
%%%%%%%%%%%%%%%%%%%%%%%%%%%%%%%%%%%%%%%%%%%%%%
\section{Approximation for larger systems}\label{Sec:LStepModel}
\subsection{General behaviour}
For systems with three or more sites, the process $\{\left(X_s,Y_s\right)\}_{s=0}^t$ is still a Markov chain with a stationary state. However, unlike the $L=2$ case, the information gained in each measurement along a trajectory $\{\Delta I_s\}_{s=0}^t$ is not a sequence of i.i.d.\ random variables. Eq.~\eqref{Eq:DeltaIwithM} cannot be reduced to a simpler form as the $M^y$ matrices for $L\geq3$ have more than one non-zero eigenvalue, and cannot in general be written down in a form that allows cancellation like~\eqref{Eq:PYcancellation}. Indeed, $\{\Delta I_s\}_{s=0}^t$ is also not a Markov chain because each value depends on the entire trajectory $\left(\mathbf{x}_s,\mathbf{y}_s\right)$ up to that point, which is a larger object at each successive value of $s$.

From simulation we that the information $I_s$ reaches linear growth after some initial transient, and so we expect the information increments $\Delta I_s$ to also converge to a stationary distribution. As the matrices used to calculate $\Delta I_s$ all have principal eigenvalue $\lambda^L_{\text{max}}<1$ (for all parameters except $1-p=r=1$), the system should exhibit an exponential decay of correlations. This allows us to assume that, at long times, $\Delta I_s$ only has significant dependence on a finite number of the previous measurement outcomes / events.

We numerically investigate the behaviour of $\Delta I_s$ for systems with $L\geq3$, specifically we focus here on results for $L=10$. Other $L$ values (larger and smaller) do not significantly differ in their general behaviour or numerics. Fig.~\ref{Fig:CDFDeltaIs} shows the CDF for $\Delta I_s$, for various system sizes up to $L=20$. The shape of the function and the position of the minimum, maximum and most likely intermediate values do not vary significantly. The scaling of the most likely, minimum and maximum of $\Delta I_s$ with $L$ is detailed in Appendices~\ref{App:TrajectoryAnalysis} and~\ref{Sec:DeltaIsBehaviour}.
\begin{figure}
\captionsetup{justification=raggedright, singlelinecheck=false}
\includegraphics[width=0.45\textwidth]{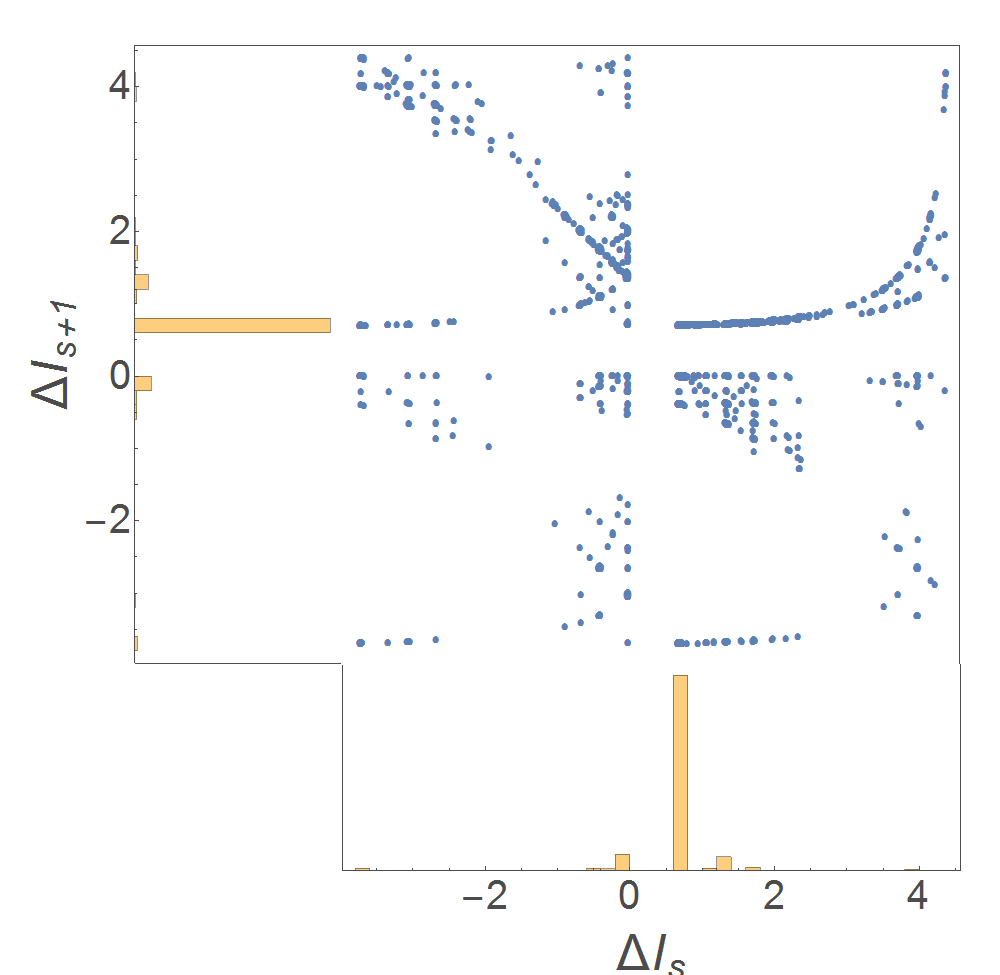}
\caption{A scatter plot of $\Delta I_{s+1}$ against $\Delta I_{s}$ for $L=10$, $p=0.5$, $r=0.9$ to illustrate correlations as explained in the text. Histograms on the axes show the density of points on the plot.}
\label{Fig:DeltaIsScatterPlot}
\end{figure}

\begin{figure}
\captionsetup{justification=raggedright, singlelinecheck=false}
\includegraphics[width=0.45\textwidth]{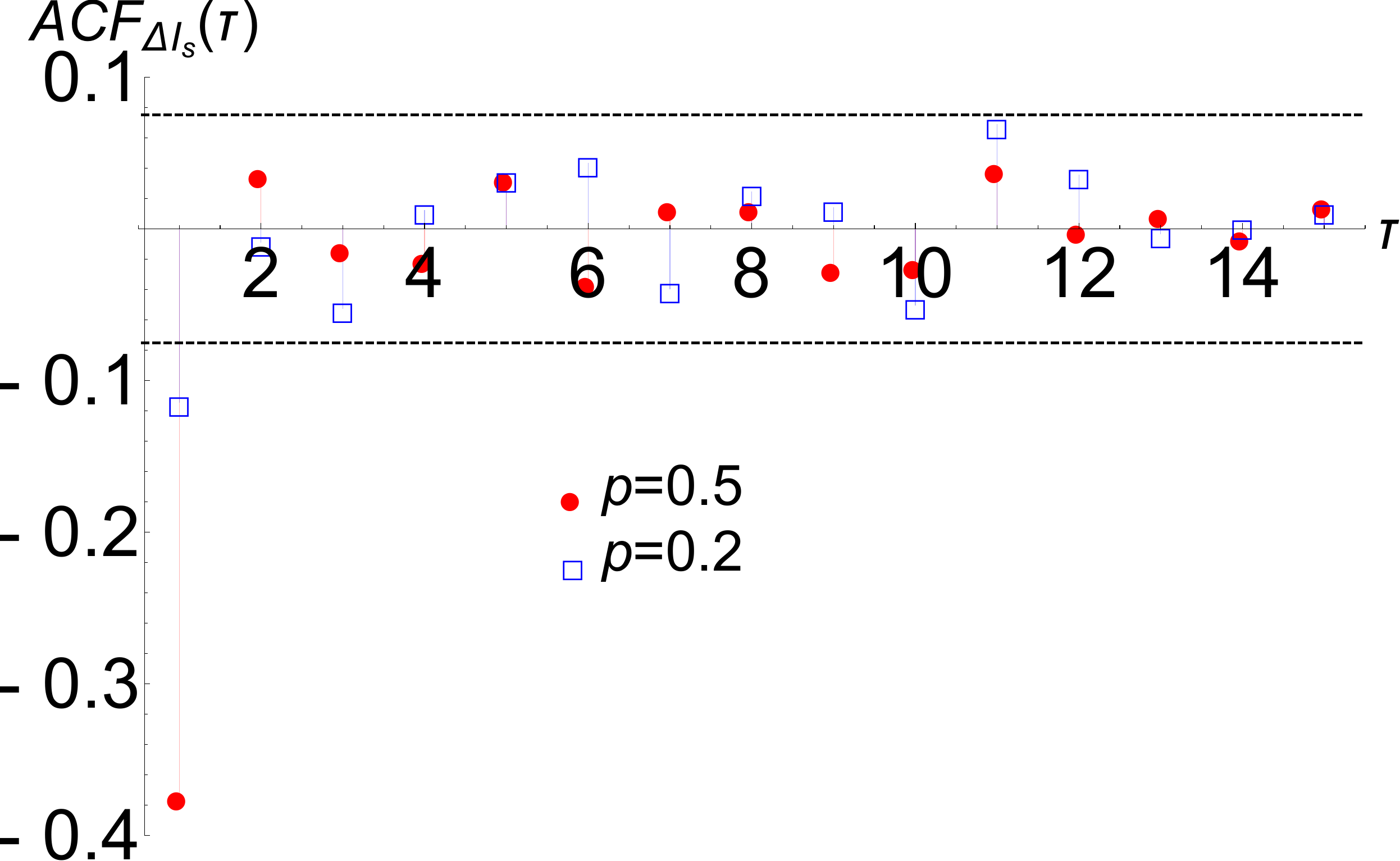}
\caption{Sample autocorrelation function of $\Delta I_s$~\eqref{Eq:ACFDef}. Shows significant negative correlations after one time step. 95\% confidence intervals plotted as blue dashed lines. $L=10$ and $r=0.9$ in both cases.}
\label{Fig:ACFPlot}
\end{figure}

Fig.~\ref{Fig:DeltaIsScatterPlot} shows a scatter plot of $\Delta I_{s+1}$ against $\Delta I_{s}$ for $L=10$. Independent random variables plotted this way would produce a symmetric cloud or grid of points. However this plot features diagonal patterns which correspond to correlation between the two variables. The projected probability densities are shown as histograms along the axes. The histograms suggest that $\Delta I_s$ and $\Delta I_{s+1}$ are identically distributed as expected. To verify this and understand how successive values of $\Delta I_s$ are correlated, we compute the sample autocorrelation function (ACF) defined as
\begin{equation}
\operatorname{\mathrm{ACF}}_{\Delta I_s}(\tau) = \frac{t\sum_{s=1}^{t-\tau}(\Delta I_{s} - \langle\Delta I\rangle)(\Delta I_{s+\tau} - \langle\Delta I\rangle)}{(t-\tau)\sum_{s=1}^{t}(\Delta I_s - \langle \Delta I\rangle)^2}.
\label{Eq:ACFDef}
\end{equation}
Fig.~\ref{Fig:ACFPlot} shows the ACF of $\Delta I$ after an initial transient period for biased and unbiased cases with 95\% white-noise confidence intervals. The autocorrelation functions show significant negative correlation between $\Delta I_s$ and $\Delta I_{s+1}$, but beyond this no significant correlation. That is, the information gained at successive time-steps is anti-correlated. This is because when an incorrect measurement is made the next measurement is likely to be correct, as correct measurements are more probable, and the amount of information gained will be positive (see App.~\ref{App:TrajectoryAnalysis}). 

While successive correct measurements do each contribute positive amounts of information, they do not differ as radically as the change from a negative to positive amount of information. Fig.~\ref{Fig:ACFPlot} also suggests that biased systems are less strongly anti-correlated than unbiased. In the next subsection we use the one time-step correlation and distribution of $\Delta I_s$ as grounds for constructing a single-step Markov chain model of $\{\Delta I_s\}_{s=0}^t$ that we believe captures most of the relevant features.
\subsection{One-step Markov chain model}
\label{Sec:1StepMarkovModel}
To obtain an approximate rate function, we assume that after an initial transient, $\{\Delta I_s\}_{s=0}^t$ is described by a stationary one-step Markov chain taking values in a continuous range. We define a finite state space $\mathcal{I}$ by coarse-graining this range and replacing $\Delta I_s$ by its expected value in each bin. The transition matrix for this process is then obtained by binning data from a scatter plot such as Fig.~\ref{Fig:DeltaIsScatterPlot} into these states and normalising the number of counts in each bin. To count the information gain, the new Markov matrix on the state space $\mathcal{I}$ is then weighted with the value of $\Delta I_s$ in the target state.
\begin{figure}
\captionsetup{justification=raggedright,singlelinecheck=false}
\includegraphics[width=0.47\textwidth]{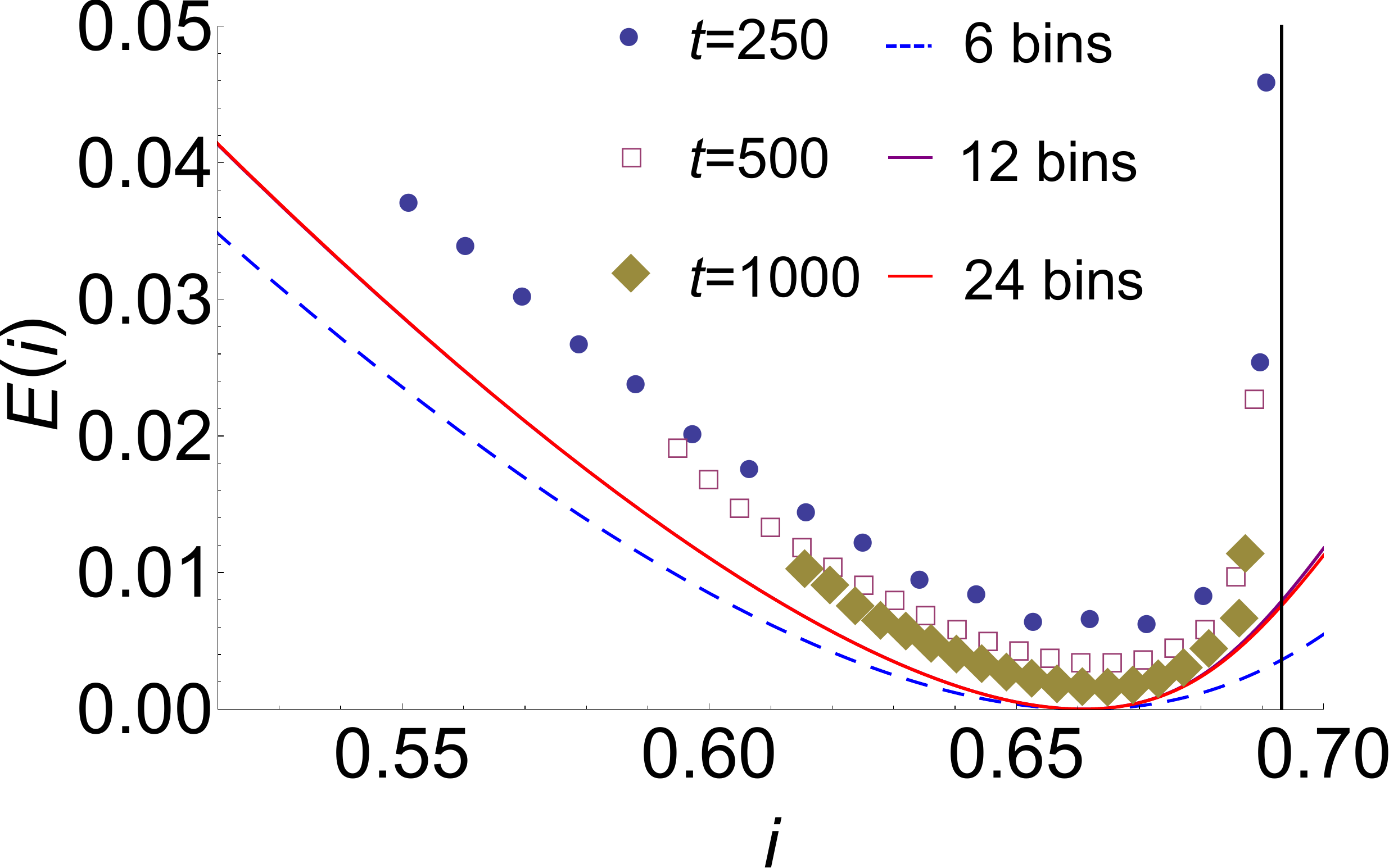}
\caption{The large deviation rate function~\eqref{Eq:LegendreFenchelTransform} for $L=10$, $p=0.5$, $r=0.9$. Lines show rate functions obtained from the Markov approximation with different numbers of bins. The lines for 12 and 24 bins are not distinguishable at this scale. Points represent data sampled at different finishing times $t$. The cutoff for positive deviations is explained in the text and given in Eq.~\eqref{Eq:CutOff}, and is represented by a vertical line.}
\label{Fig:LSiteRateFn}
\end{figure}
The SCGF (and thus the rate function) can then be obtained from the largest eigenvalue of this tilted transition matrix. 

To check the method it can be shown that for $L=2$, as the number of bins is increased and more data is used in the scatter plot, the method converges to the analytically obtained rate function for that case. Fig.~\ref{Fig:LSiteRateFn} shows the rate function obtained through this method for $L=10$ plotted alongside data obtained from simulation. The figure shows convergence of the estimated rate functions with increasing number of bins, which appears to be consistent with the data. This confirms the validity of the one-step Markov approximation for a wide range of fluctuations. However, as can be seen in Fig.~\ref{Fig:LSiteRateFn}, the data indicate a cut-off in the rate function that the Markov approximation does not predict. In the next subsection we explain this feature by noting that large fluctuations of $\Delta I_s$ do not accumulate in the way that the Markov model allows.
\subsection{Beyond Markovian analysis}
\label{Sec:BeyondMarkov}
To correct the numerically obtained rate functions shown by solid lines in Fig.~\ref{Fig:LSiteRateFn} we must consider the maximum possible value for $I_t / t$. Fig.~\ref{Fig:DeltaIsScatterPlot} suggests that it is possible to obtain large amounts of information on consecutive time-steps (the top right corner of this scatter-plot has a small but non-zero population). However, investigation of individual trajectories of the system reveals that consecutive large positive amounts can only be obtained after consecutive large negative amounts (see App.~\ref{App:TrajectoryAnalysis}). This is not reflected in the ACF in Fig.~\ref{Fig:ACFPlot} owing to the fact that these events occur very rarely, as can be seen from the marginal histograms in Fig.~\ref{Fig:DeltaIsScatterPlot}.

The maximum value of $I_t / t$ is obtained by measuring correctly every time-step. This maximum value is given by
\begin{equation}
\frac{I^u_\mathrm{max}}{t} = \ln{\frac{r}{\alpha}},
\label{Eq:CutOff}
\end{equation}
where $\alpha$ is numerically obtained from the $M_y$ matrices for that system. A discussion is included in App.~\ref{App:TrajectoryAnalysis}. 

Fig.~\ref{Fig:LSiteRateFn} shows the cut-off value for an unbiased system with a black vertical line. Unlikely trajectories in biased systems that always step against the bias can generate large positive deviations of $I_t / t$ and so the rate function cuts off at higher values. The cut-off is therefore less relevant when comparing data for biased systems to the predicted rate function.

The $\Delta I_s$ process is clearly not a one-step Markov chain, and the rate function obtained this way is also limited by finite sampling of transitions and limitations on the number of bins used. However, the one-step Markov model gives a rate function that converges reasonably quickly with increasing number of data points and together with the cut-off, captures well the shape of the sampled data in a way that a Gaussian or i.i.d.\ approximation would not. An $n$-step Markov chain model might also capture this but it appears that simply including the cut-off at the maximum value is sufficient to obtain a good approximate rate function.
%%%%%%%%%%%%%%%%%%%%%%%%%%%%%%%%%%%%%%%%%%%%%%
%%%%%%%%%%%%%%%%%%%%%%%%%%%%%%%%%%%%%%%%%%%%%%
\section{Discussion}\label{Sec:Discussion}

The information gain $I_t$ is a quantity recently introduced in the analysis of feedback systems~\cite{parrondo2015thermodynamics,horowitz2010nonequilibrium} and studied as a component in the development of information thermodynamics and information engines~\cite{horowitz2011thermodynamic}. The fluctuation properties of this quantity are relevant when considering information processing in feedback devices; the quantity of information gained is directly proportional to the work required to delete that information from the feedback device's memory. 

In this paper we have studied a simple model of an information engine and obtained an exact analytical expression of the large deviation rate function for information gain $I_t$ in a two-site system. For larger systems we have shown that a one-step Markov approximation captures most of the relevant details of the large deviations. We are also able to predict the cut-off of this rate function by considering the maximum amount of information that can be obtained.

Significantly, the one-step Markov approximation allows us to easily obtain an approximation of the rate function from data by sampling the information gain at consecutive time steps. This is computationally easier than directly sampling the distribution of $I_t/t$ (which requires very long times or cloning-type algorithms \cite{rohwer2014convergence}) but together with the theoretically predicted cut-off seems to provide a consistent estimate of the shape of the large deviation rate function.

The rate functions obtained here demonstrate that the information gained by the measuring device in a simple Markovian feedback system shows a strong asymmetry around the mean. The cut-off value for this rate function is sensitive to the dynamics of the system, namely whether the particle motion is symmetric or asymmetric. It would be of interest to study other information engines to check whether these findings are generic and to what extent the Markov approximation is applicable in other systems. The large deviation rate function offers the possibility to explore detailed fluctuation relationships beyond~\eqref{Eq:GIFR} for $I_t$. To obtain a detailed fluctuation relationship, care must be taken in deciding how to meaningfully time-reverse a feedback system. Discussions have already alluded~\cite{ponmurugan2010generalized,horowitz2010nonequilibrium} to the potential difficulties in interpreting time-reversed feedback in a physically meaningful manner and there is evidently much scope for future work.

%%%%%%%%%%%%%%%%%%%%%%%%%%%%%%%%%%%%%%%%%%%%%%
%%%%%%%%%%%%%%%%%%%%%%%%%%%%%%%%%%%%%%%%%%%%%%
\section*{Acknowledgements}
This work was supported by the Engineering and Physical Sciences Research Council (EPSRC), Grant No.\ EP/I01358X/1. We would also like to thank Takahiro Sagawa, Sosuke Ito, Jordan Horowitz and Hugo Touchette for helpful discussions. RJH is grateful for the hospitality of the National Institute for Theoretical Physics (NITheP) Stellenbosch.

%%%%%%%%%%%%%%%%%%%%%%%%%%%%%%%%%%%%%%%%%%%%%%
%%%%%%%%%%%%%%%%%%%%%%%%%%%%%%%%%%%%%%%%%%%%%%
\appendix

\section{Trajectory Analysis}
\label{App:TrajectoryAnalysis}

To understand the various events that occur in the system, we plot trajectories of $\Delta I_s$, $P(\mathbf{y}_{s-1})/P(\mathbf{y}_s)$, $X_s$ and $Y_s$. Recall that $\Delta I_s$ is given by Eq.~\eqref{Eq:DeltaIDef} (reproduced here for readability)
\begin{equation}
\Delta I_s = \ln{p(y_s\mid x_s)}+\ln{\frac{P(\mathbf{y}_{s-1})}{P(\mathbf{y}_s)}},\nonumber
\end{equation}
and hence $\Delta I_s$ will always differ from $\ln{P(\mathbf{y}_{s-1})/P(\mathbf{y}_s)}$ by $\ln{r}$ or $\ln{w}$ depending on whether the measurement is correct or incorrect.

The value of $P(\mathbf{y}_{s-1})/P(\mathbf{y}_s)$ is determined by the measurement trajectory itself and is not directly dependent on the trajectory $\mathbf{x}_t$. We observe that there are two `baseline' values for this ratio that are obtained when the measurements represent a possible trajectory for the particle, i.e.\ the particle does not appear to jump more than one site in a single time-step and does not appear to move `through' the barrier. These two baseline values correspond to the particle jumping up or being blocked attempting to jump down.

If the particle is blocked at site $x$ for successive steps starting, for example, at time $s-3$, the probability of the correct measurement history is calculated via a matrix product as in Eq.~\eqref{Eq:PYwithM},
\begin{equation}
P(\mathbf{y}_{s}) = \bra{v_{s-3}} M_{y}M_{y}M_{y} \ket{1},
\label{Eq:PYBlockedJumps}
\end{equation}
where $y=x$ and where $\bra{v_{s-3}}=\bra{\pi_0}\prod_{k=0}^{s-3} M_{y_{k}}$. The ratio $P(\mathbf{y}_{s-1})/P(\mathbf{y}_{s})$ entering the information gain $\Delta I_s$ (Eq.~\eqref{Eq:DeltaIwithM}) is then dominated by the leading eigenvalue $\lambda_{\mathrm{max}}^L$ of the matrices, and therefore should approach $1/\lambda^L_\mathrm{max}$ very quickly. Recall that all matrices have the same eigenvalues since they are related by translations of rows and columns. The corresponding lower baseline value for $\Delta I_s$ is given by
\begin{equation}
\Delta I^b = \ln{\frac{r}{\lambda^L_{\mathrm{max}}}}.
\label{Eq:LowerBaseLine}
\end{equation}

On the other hand, let us assume that the particle jumps up for successive time steps, starting in site $x$ at time $s-3$, and we measure this correctly. Then the probability of the measurement history $P(\mathbf{y}_s)$ is given by a product of matrices with increasing index
\begin{equation}
P(\mathbf{y}_{s}) = \bra{v_{s-3}} M_{y}M_{y+1}M_{y+2} \ket{1},
\label{Eq:PYUpJumps}
\end{equation}
where $y=x$, and $y+n$ is understood with periodic boundary conditions. The successive matrices are translated by one column and one row, i.e.\ $M_{y+1}=T M_y T^{-1}$, where the translation $T$ is such that
\begin{equation}
\left(\bra{v}T\right)_k = v_{k+1} ~\text{and}~\left(T\ket{w}\right)_k=w_{k-1}.
\end{equation}
Similar to the eigenvalue case, the expression~\eqref{Eq:PYUpJumps} then is dominated by vectors $\bra{v}$ and a scale factor $\alpha$ such that
\begin{equation}
\left(\bra{v} M_y\right)_k = \alpha v_{k+1},
\label{Eq:GetAlpha}
\end{equation}
with periodic boundaries (i.e. $\left(\bra{v} M_y\right)_1= \alpha v_L$). So the upper baseline value for information gain $\Delta I_s$ should be given by
\begin{equation}
\Delta I^u = \ln{\frac{r}{\alpha}},
\label{Eq:UpperBaseLine}
\end{equation}
where $\alpha$ can be found numerically from~\eqref{Eq:GetAlpha} for any given system.

\begin{figure}
\captionsetup{justification=raggedright, singlelinecheck=false}
\includegraphics[width=0.45\textwidth]{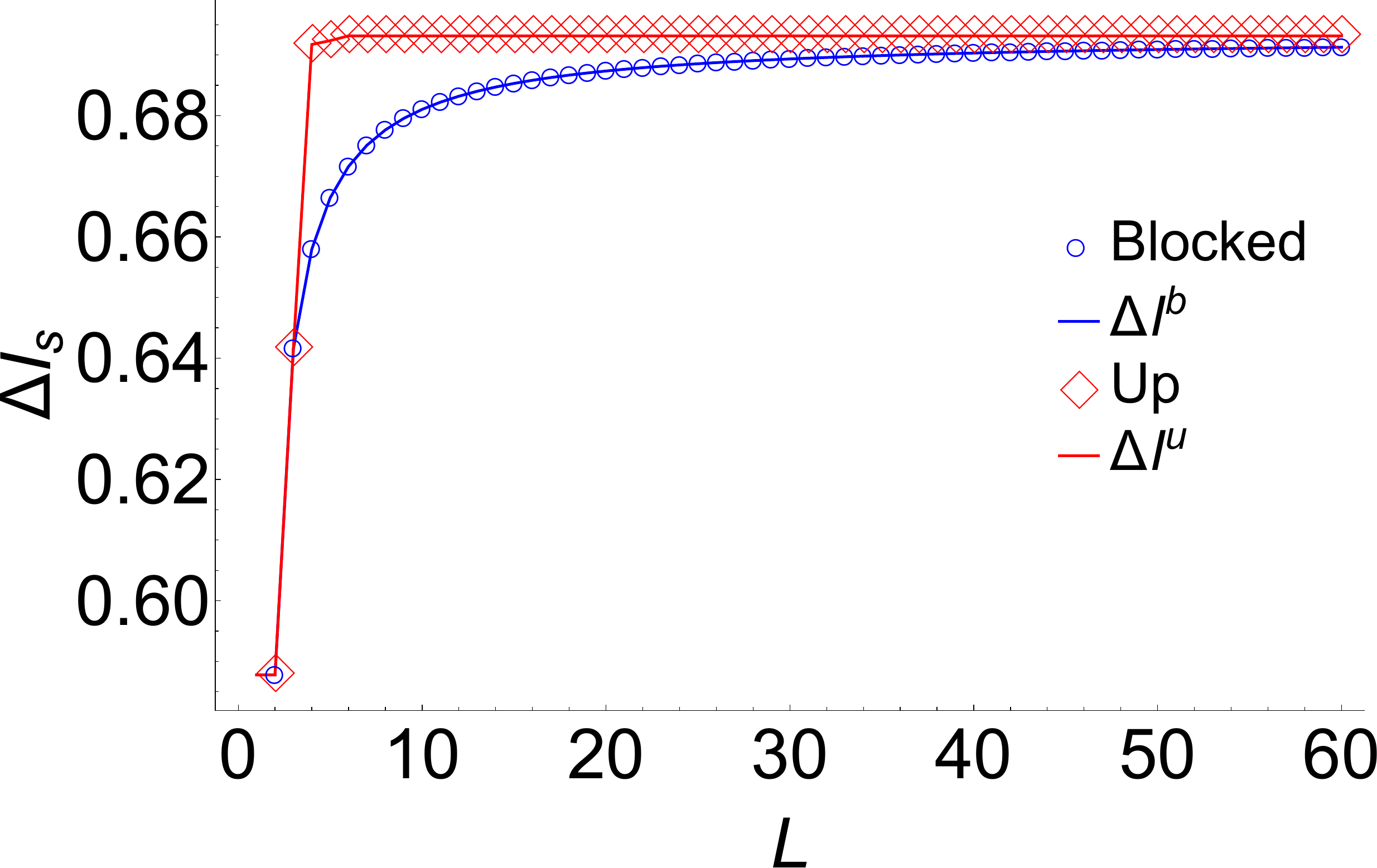}
\caption{Baseline values for an unbiased system ($p=q$) for $r=0.9$. Points show the value of $\Delta I_s$ for consecutive up jumps and blocked down jumps. Lines show the predictions from~\eqref{Eq:LowerBaseLine} and~\eqref{Eq:UpperBaseLine}.}
\label{Fig:PRatioUnbiased}
\end{figure}

\begin{figure}
\captionsetup{justification=raggedright, singlelinecheck=false}
\includegraphics[width=0.45\textwidth]{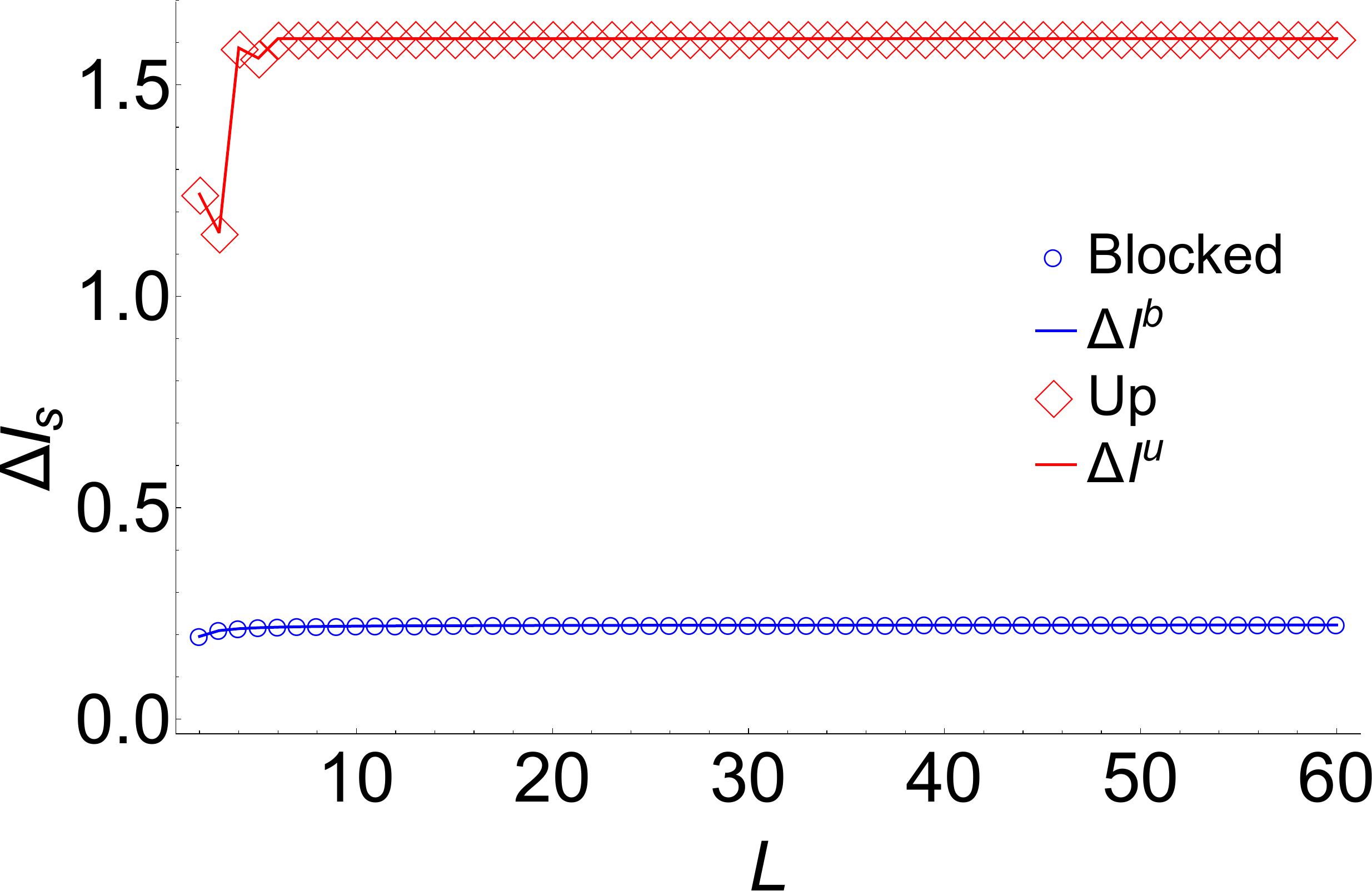}
\caption{Baseline values for a biased system, $p=0.2$ and $r=0.9$. Points show the value of $\Delta I_s$ for consecutive up jumps and blocked down jumps. Lines show the predictions from~\eqref{Eq:LowerBaseLine} and~\eqref{Eq:UpperBaseLine}.}
\label{Fig:PRatioBiased}
\end{figure}

\begin{figure}
\captionsetup{justification=raggedright, singlelinecheck=false}
\includegraphics[width=0.45\textwidth]{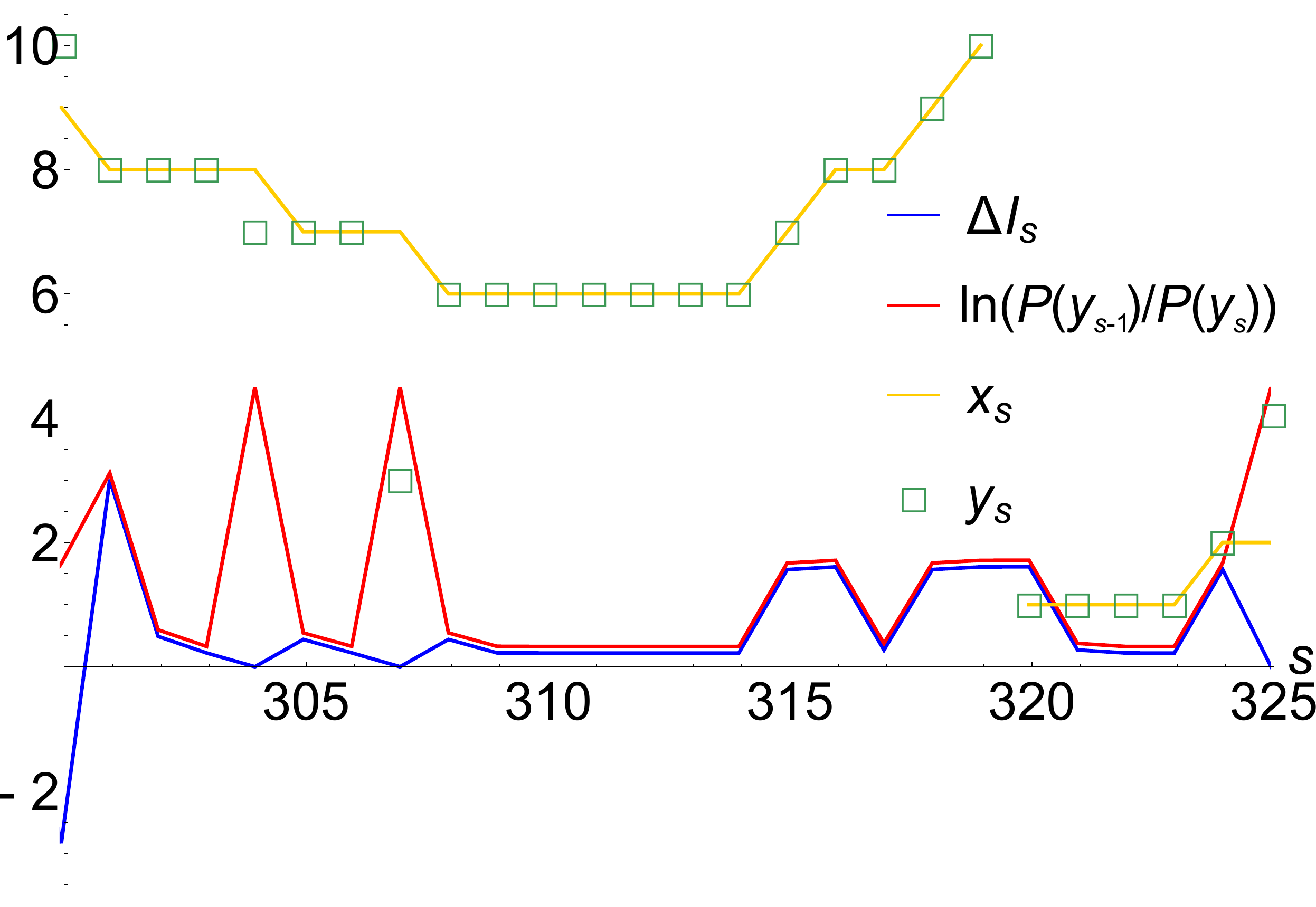}
\caption{Trajectory of a biased system for $L=10$, $p=0.2$, $r=0.9$. The trajectory $x_s$ is given by a gold line, and measurements by square boxes. The information gain $\Delta I_s$, given by a blue line, is always bounded above by $\ln(P(\mathbf{y}_{s-1})/P(\mathbf{y}_s))$ given by a red line. Note that these quantities have different units. Note also that for $x_s$ and $y_s$ there are periodic boundary conditions between 0 and 10.}
\label{Fig:AsymmetricJumpUp}
\end{figure}

Figs.~\ref{Fig:PRatioUnbiased} and~\ref{Fig:PRatioBiased} show empirical data confirming our predictions for the baseline values and demonstrate their scaling with $L$ in unbiased and biased systems, respectively. In an unbiased system, the difference in $P(\mathbf{y}_{s-1})/P(\mathbf{y}_s)$ between an up and blocked down jump shrinks with increasing $L$, whereas the values are roughly constant for biased systems. Fig.~\ref{Fig:AsymmetricJumpUp} shows a typical section of a trajectory in a biased system. The baseline values of $P(\mathbf{y}_{s-1})/P(\mathbf{y}_s)$ and $\Delta I_s$ for blocked down and up jumps are seen around $t=310$ and $t=315$ respectively.

\begin{figure}
\captionsetup{justification=raggedright, singlelinecheck=false}
\includegraphics[width=0.47\textwidth]{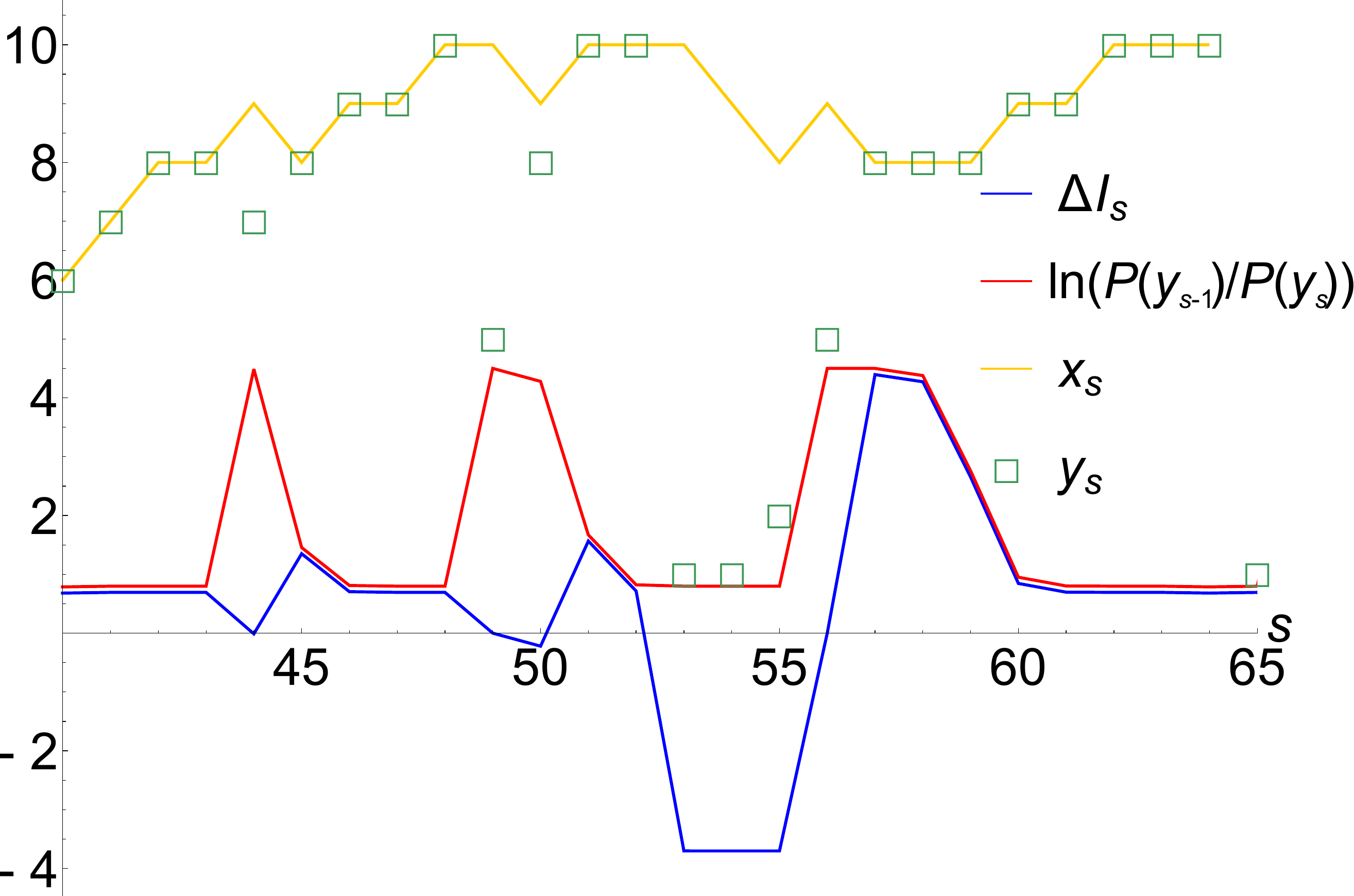}
\caption{Trajectory of a biased system for $L=10$, $p=0.5$, $r=0.9$. See Fig.~\ref{Fig:AsymmetricJumpUp} for details of the plot.}
\label{Fig:ConsecutiveIncorrectMeasurements}
\end{figure}

Whenever a measurement is made that is incompatible with previous measurements (i.e.\ the particle looks to have jumped two or more sites or moved through the barrier), $P(\mathbf{y}_{s-1})/P(\mathbf{y}_s)$ changes value. Isolated incorrect measurements as seen in Fig.~\ref{Fig:AsymmetricJumpUp} (at time $t=307$) and Fig.~\ref{Fig:ConsecutiveIncorrectMeasurements} (at time $t=44$) cause $P(\mathbf{y}_{s-1})/P(\mathbf{y}_s)$ to increase while we observe that the $\ln{w}$ term means that $\Delta I_s=0$. In the following measurements, $P(\mathbf{y}_{s-1})/P(\mathbf{y}_s)$ is still larger than its baseline value and as the $\ln{r}$ contribution is small in comparison, $\Delta I_s$ is also larger than the baseline.

The information theoretic interpretation of these observations is that upon making a measurement that is not compatible with the previous measurements, no new information is gained. This is because it is not clear to the observer whether the current measurement is incorrect, or the previous measurements were erroneous (or both). It is only on subsequent measurements that information is gained, as more measurements allow the observer to make inferences about which measurements were incorrect. After an incorrect measurement, $P(\mathbf{y}_{s-1})/P(\mathbf{y}_s)$ (and hence $\Delta I_s$) returns quickly to a baseline value.

To observe very large values of $\Delta I_s$ and instantaneously gain large amounts of information, it is necessary to first lose larger amounts of information through incorrect measurements. An example of this is shown in Fig.~\ref{Fig:ConsecutiveIncorrectMeasurements}, where the large amounts of information gained between $t=57$ and $t=59$ cannot balance the losses between $t=53$ and $t=55$. Hence it is not possible to gain additional information by making strategically `wrong' measurements, as a series of correct measurements would yield more total information.

\begin{figure}
\captionsetup{justification=raggedright, singlelinecheck=false}
\includegraphics[width=0.45\textwidth]{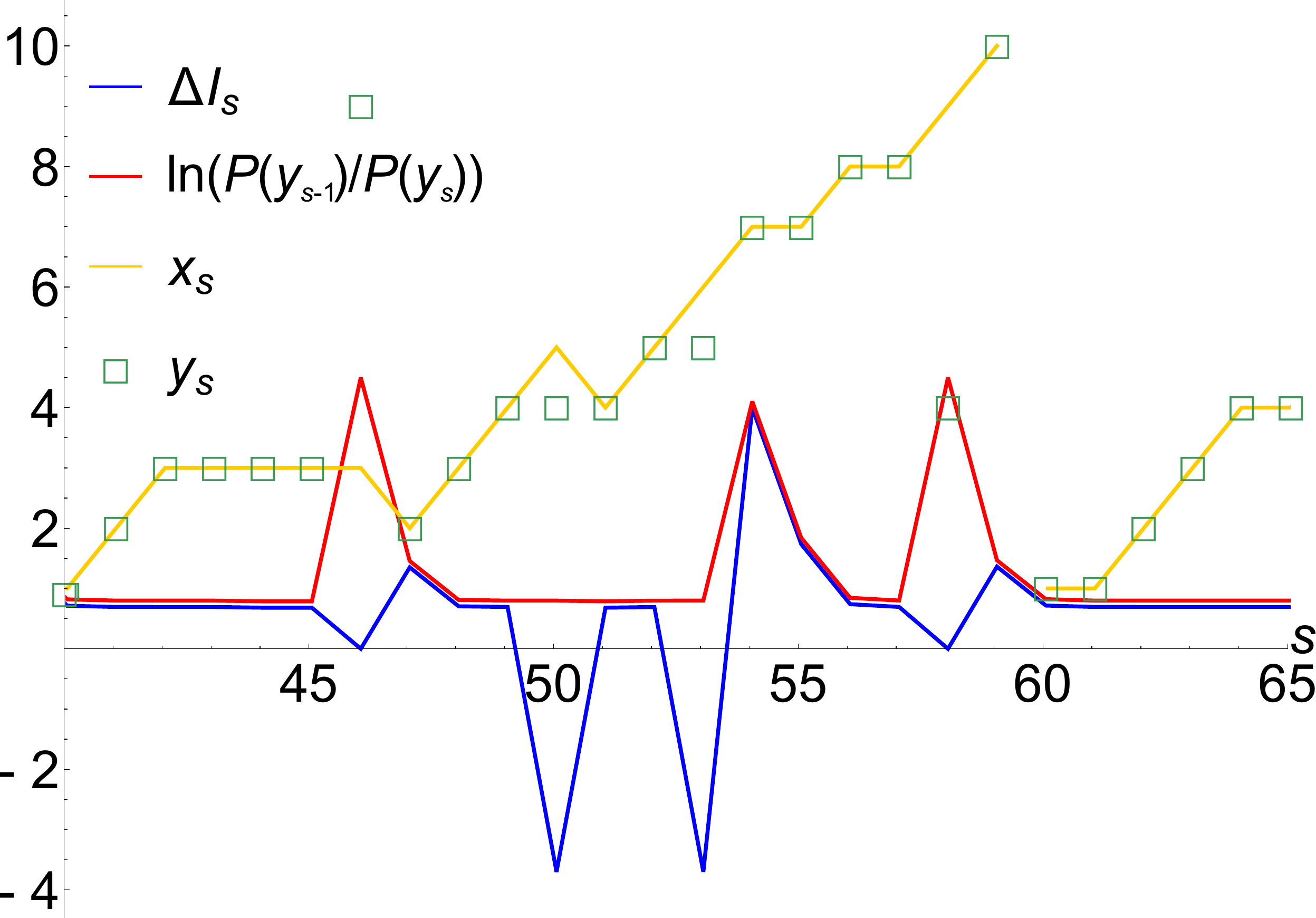}
\caption{Trajectory of an unbiased system for $L=10$, $p=0.5$, $r=0.9$. See Fig.~\ref{Fig:AsymmetricJumpUp} for details of the plot.}
\label{Fig:UndetectableIncorrectMeasurements}
\end{figure}

In the case of a wrong measurement that still represents a possible trajectory for the particle, $P(\mathbf{y}_{s-1})/P(\mathbf{y}_s)$ does not change but the $\ln{w}$ contribution from the incorrect measurement means that $\Delta I_s$ takes a negative value. If the following measurements are correct and also compatible with the previous wrong measurement, then subsequent measurements will only gain a baseline amount of information. An example of this is shown in Fig.~\ref{Fig:UndetectableIncorrectMeasurements}. Here, a wrong measurement occurs at time $t=50$ which is compatible with the previous history. Subsequent measurements do not allow the observer to ascertain that any of previous measurements were incorrect and hence this information loss is not recovered.
%%%%%%%%%%%%%%%%%%%%%%%%%%%%%%%%%%%%%%%%%%%%%%
%%%%%%%%%%%%%%%%%%%%%%%%%%%%%%%%%%%%%%%%%%%%%%

\section{Independence of $\Delta I_s$ for $L=2$}\label{App:IID}

To demonstrate that $\Delta I_s$ are i.i.d.\ random variables for $L=2$, we want to show that the distribution of $\Delta I_{s+1}$ is independent of $\Delta I_s$ and identical for all $s$.  Let us first assume that the system is in the state $(x_s,y_s)=(1,1)$. The probabilities for $\Delta I_{s+1}$ taking the values in~\eqref{Eq:2SiteDeltaIsA} and \eqref{Eq:2SiteDeltaIsD} are as follows:
\begin{center}
\begin{tabular}{| c | c | c |}
\hline
 $\Delta I_{s+1}$ & Transition & Probability \\ \hline
$a$ & $(1,1)\to(2,2)$ & $p r$ \\ \hline
$b$ & $(1,1)\to(1,1)$ & $q r$ \\ \hline
$c$ & $(1,1)\to(1,2)$ & $q w$ \\ \hline
$d$ & $(1,1)\to(2,1)$ & $p w$ \\ \hline
\end{tabular}
\end{center}
By the translation invariance in the system, the same probabilities apply if the system was in state $(x_s,y_s)=(2,2)$.

If the system starts in the state $(x_s,y_s)=(1,2)$, then the probabilities are:
\begin{center}
\begin{tabular}{| c | c | c |}
\hline
 $\Delta I_{s+1}$ & Transition & Probability \\ \hline
$a$ & $(1,2)\to(1,1)$ & $p r$ \\ \hline
$b$& $(1,2)\to(2,2)$ &$q r$ \\ \hline
$c$ & $(1,2)\to(2,1)$& $q w$ \\ \hline
$d$ & $(1,2)\to(1,2)$ & $p w$ \\ \hline
\end{tabular}
\end{center}
which again by translation invariance also holds for beginning in the state $(2,1)$. We can see then that the probability to obtain $\Delta I_{s+1}$ is in fact independent of the state $\left(x_s,y_s\right)$ of the system. In particular, it does not depend on the previous value $\Delta I_s$. Therefore, $\{\Delta I_s\}_{s=0}^t$ is indeed a sequence of i.i.d.\ random variables.

%%%%%%%%%%%%%%%%%%%%%%%%%%%%%%%%%%%%%%%%%%%%%%
%%%%%%%%%%%%%%%%%%%%%%%%%%%%%%%%%%%%%%%%%%%%%%
\section{Maximum and Minimum of $\Delta I_s$}
\label{Sec:DeltaIsBehaviour}

\begin{figure}
\captionsetup{justification=raggedright, singlelinecheck=false}
\includegraphics[width=0.45\textwidth]{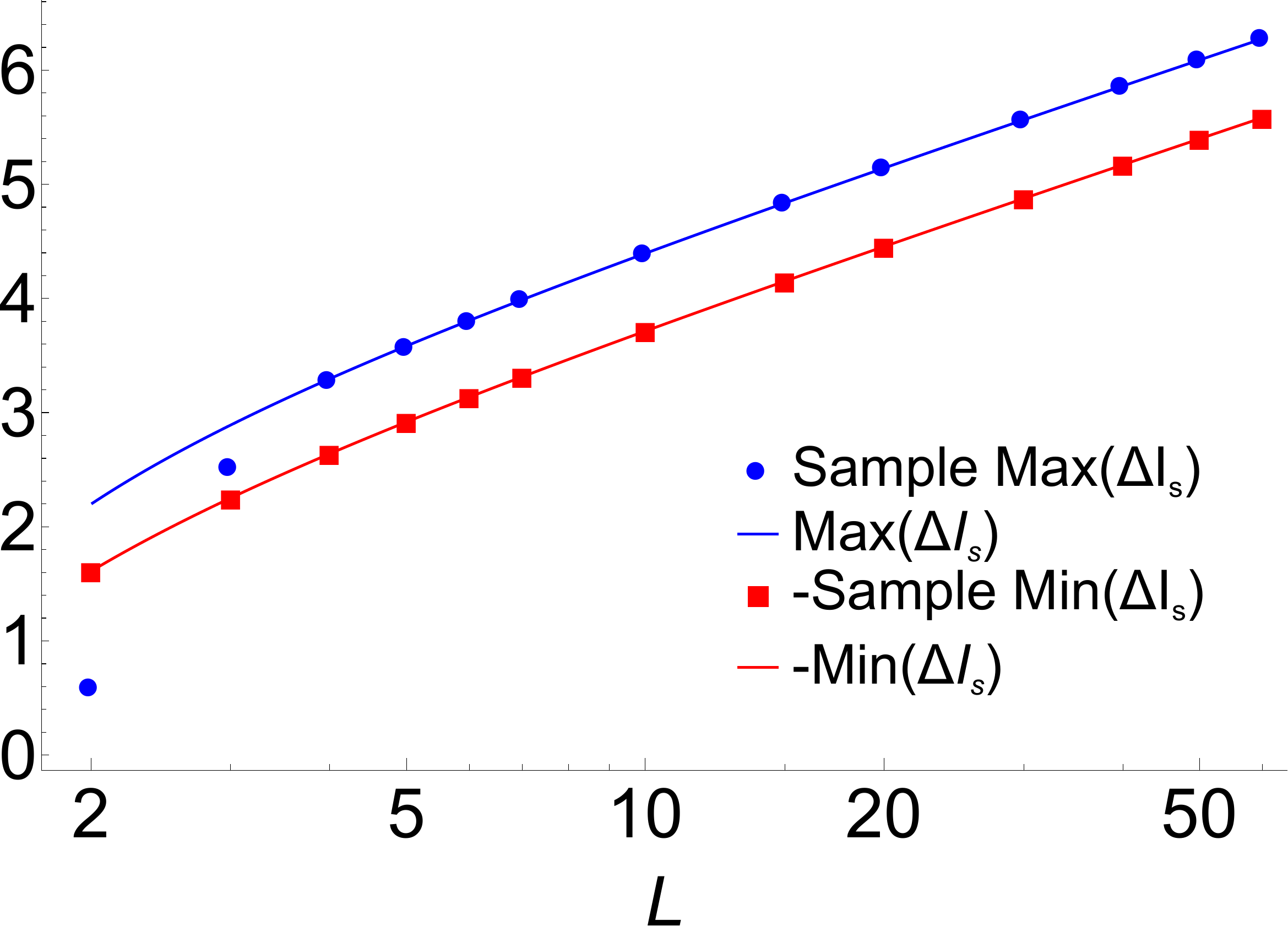}
\caption{Points show numerical results for $\operatorname{\mathrm{Max}}(\Delta I_s)$ and $-\operatorname{\mathrm{Min}}(\Delta I_s)$ (from $10^7$ trajectory realisations of length $t=500$) and predicted values given by~\eqref{Eq:MinDeltaIs} and~\eqref{Eq:MaxDeltaIs} for varying $L$ for $p=0.5$, $r=0.9$.}
\label{Fig:MaxMinDeltaIsPlot}
\end{figure}

Fig.~\ref{Fig:MaxMinDeltaIsPlot} shows numerical results for the maximum and minimum of $\Delta I_s$ from $10^7$ realisations up to $t=1000$ for varying $L$. The minimum amount of information per time-step is obtained when an incorrect measurement is made and we argue that, as in the $L=2$ case (see Eq.~\eqref{Eq:2SiteDeltaIsC}) its value is given by
\begin{equation}
\operatorname{\mathrm{Min}}(\Delta I_s) = \ln{\frac{w}{\lambda^L_\mathrm{max}}}.
\label{Eq:MinDeltaIs}
\end{equation}
The maximum value is observed to be exactly the minimum value reflected across the lower baseline value for information gain~\eqref{Eq:LowerBaseLine}, i.e.\ it is given by, 
\begin{equation}
\operatorname{\mathrm{Max}}(\Delta I_s) = \ln{\frac{r}{\lambda^L_\mathrm{max}}}  -\operatorname{\mathrm{Min}}(\Delta I_s) = \ln{\frac{r}{w}}.
\label{Eq:MaxDeltaIs}
\end{equation}
This holds for all $L\geq4$ as shown by the numerical results in Fig.~\ref{Fig:MaxMinDeltaIsPlot}. For $L\leq3$, the observed discrepancy is probably due to the fact that incorrect measurements are more constrained e.g.\ any barrier placement in a two-site system will interfere with the particle motion.

%%%%%%%%%%%%%%%%%%%%%%%%%%%%%%%%%%%%%%%%%%%%%%
%%%%%%%%%%%%%%%%%%%%%%%%%%%%%%%%%%%%%%%%%%%%%%

%\bibliography{references}


\end{document}